\documentclass[10pt,a4paper,twocolumn]{article}
\pdfoutput=1
\usepackage[utf8]{inputenc}
\usepackage{amsmath}
\usepackage{amssymb}
\usepackage{graphicx}
\usepackage{geometry}
\usepackage{physics}
\usepackage{cite}

\title{Quantum interference between photons from an atomic ensemble and a remote atomic ion}
\author{A.N. Craddock${}^{1,\dagger}$, J. Hannegan${}^{1,\dagger}$, D.P. Ornelas-Huerta${}^{1}$, J.D. Siverns${}^{1}$, A.J. Hachtel${}^{1}$, \\ E.A. Goldschmidt${}^{2}$, J.V. Porto${}^{1}$, Q. Quraishi${}^{1,2}$, S.L. Rolston${}^{1,*}$\\
{\small ${}^{1}$Joint Quantum Institute, National Institute of Standards and Technology and the University of Maryland,}\\
{\small College Park, Maryland 20742 USA}\\
{\small ${}^{2}$Army Research Laboratory, 2800 Powder Mill Rd., Adelphi, MD 20783}\\
{\small ${}^{\dagger}$These authors contributed equally}\\
{\small ${}^{*}$rolston@umd.edu}}

\begin{document}

\maketitle

\textbf{Advances in the distribution of quantum information will likely require entanglement shared across a hybrid quantum network \cite{photonicstatetransfer,Lettner2011,Meyer2015}.
Many entanglement protocols require the generation of indistinguishable photons between the various nodes of the network~\cite{Sangouard2011,simon2003robustentanglement}.
This is challenging in a hybrid environment due to typically large differences in the spectral and temporal characteristics of single photons generated in different systems~\cite{photonicstatetransfer}.
Here we show, for the first time, quantum interference between photons generated from a single atomic ion and an atomic ensemble, located in different buildings and linked via optical fibre.
Trapped ions are leading candidates for quantum computation and simulation with good matter-to-photon conversion~\cite{figgatt2017complete,haffner2008quantum,blatt2012quantum,zhang2017ionphasetransistion,longionmemory,harty2014high,hucul2015modular,highfidelityionphotontelecom}.
Rydberg excitations in neutral-atom ensembles show great promise as interfaces for the storage and manipulation of photonic qubits with excellent efficiencies~\cite{Lampen2018,Dudin2012,Thompson2017,Gorniaczyk2014}.
Our measurement of high-visibility interference between photons generated by these two,
disparate systems is an important building block for the establishment of a hybrid quantum network.}

Recently, Rydberg atoms have proven to be a useful tool in the field of quantum information. 
The strong optical nonlinearity exhibited by neutral-atom Rydberg ensembles enables the construction of single-photon sources~\cite{Dudin2012}, gates~\cite{Thompson2017}, and transistors~\cite{Gorniaczyk2014}. 
Strong light-matter interactions make them well suited as quantum memories~\cite{Lampen2018}, and for implementing quantum repeaters~\cite{Solmeyer2016,Zhao2010}. 
Furthermore, arrays of Rydberg atoms are a powerful new platform for quantum simulation~\cite{DeLeseleuc2017,Bernien2017}. 
The continued success of trapped-ion systems in quantum computation\cite{figgatt2017complete,haffner2008quantum}, simulation\cite{blatt2012quantum,zhang2017ionphasetransistion}, and communication \cite{hucul2015modular} owes to their long coherence and trapping lifetimes \cite{longionmemory}, high fidelity operations \cite{harty2014high}, and ease of generating ion-photon entanglement \cite{hucul2015modular,highfidelityionphotontelecom}.

Given the wide-ranging applications of both platforms, future efforts in quantum information will benefit from the construction of remote hybrid atomic-ensemble-ion networks.
Flying photonic qubits provide an excellent means of connecting nodes~\cite{Cirac1997,Olmschenk}.
Many entangling protocols require some degree of indistinguishability between photons produced by the different network nodes~\cite{Sangouard2011,simon2003robustentanglement}, however, photons generated by different atomic systems are in general spectrally distinguishable.
Although this is not a physical limitation~\cite{legero2003,Olmschenk}, the necessity for detectors with bandwidths orders-of-magnitude greater than currently available, along with vanishing entanglement generation rates has prohibited the linking of such systems.
In this work, we achieve high-visibility Hong-Ou-Mandel (HOM) interference~\cite{Hong1987}, indicative of a high degree of indistinguishability, between photons generated from a rubidium atomic ensemble and a trapped barium ion after closely matching their centre frequencies via difference frequency generation (DFG)~\cite{siverns2018neutral}.
Furthermore, we analyse the feasibility of hybrid ion-Rydberg remote entanglement generation.

\begin{figure*}
    \centering
    \includegraphics[width=1\textwidth]{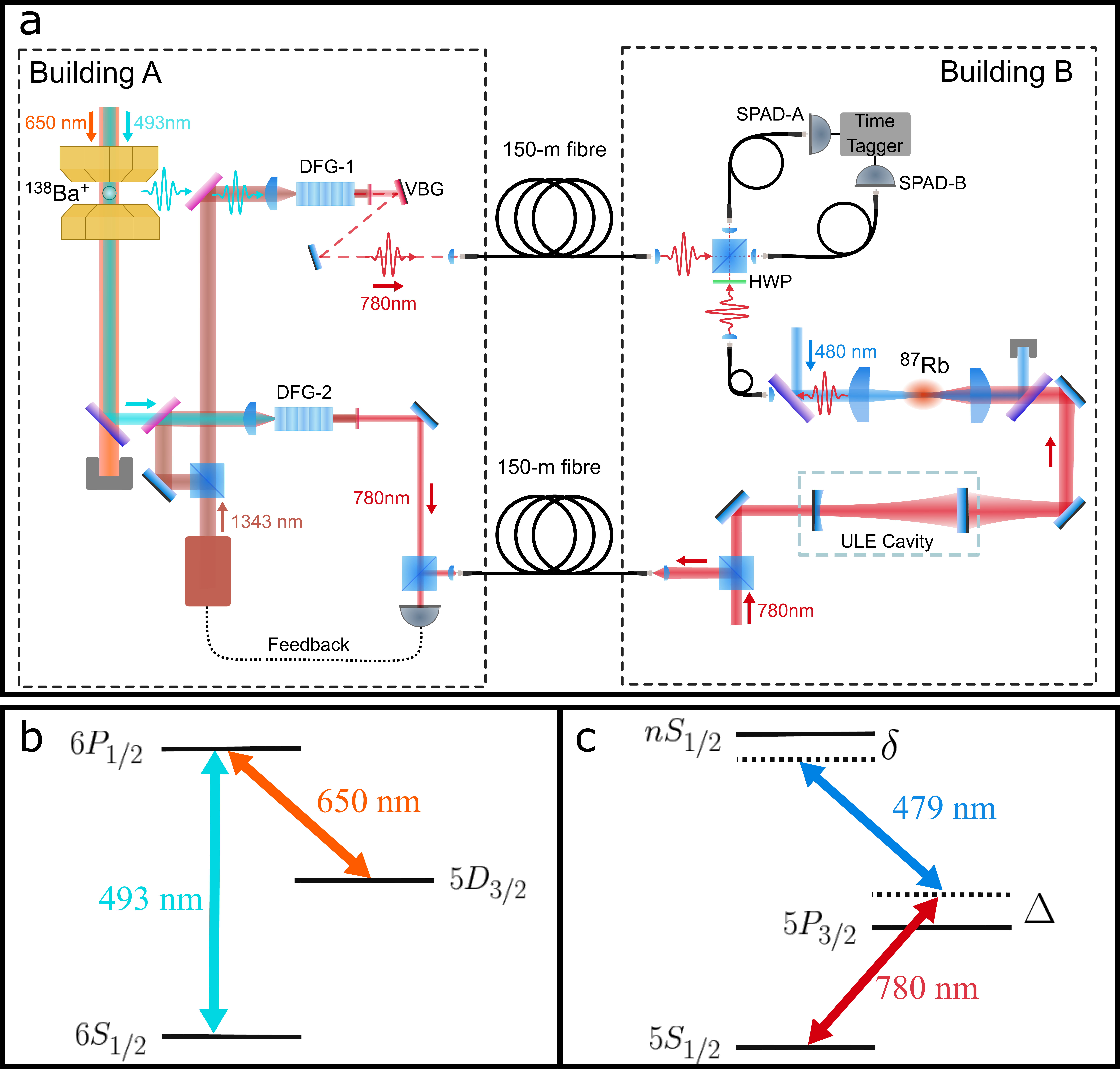}
    \caption{\textbf{Experimental layout and energy level diagrams for the two sources.} \ \textbf{a}, Building A contains a $^{138}$Ba$^{+}$ ion which emits photons at 493-nm, and Building B contains a $^{87}$Rb atomic ensemble producing 780-nm photons. 
    Ion-emitted photons are converted to 780-nm using DFG-1 and sent to Building B via PMF. 
    DFG-2 produces 780-nm light used to frequency stabilise the output of DFG-1 by optical beatnote locking with reference light sent from Building B. 
    Light from the ion and ensemble source is sent to the HOM interferometer for two-photon interference measurements. 
    A half-wave plate (HWP) in one input path allows for control of the relative polarisation of the photons.
    The photons interfere on a nearly 50:50 beamsplitter before being coupled into two SMF which are connected to SPADs linked to a time-tagging device.
    Here VBG is a volume Bragg grating and the ULE cavity is an ultra-low expansion cavity.
    \textbf{b}, Level scheme for $^{138}$Ba$^{+}$. 
    \textbf{c}, Level scheme for $^{87}$Rb. 
    }
    \label{fig:layout}
\end{figure*}

Our experiment spans two buildings, shown in Fig.~\ref{fig:layout}. 
Building A contains a single trapped $^{138}$Ba$^+$ ion as well as two DFG setups.
Building B contains a $^{87}$Rb atomic ensemble and a HOM interferometer to measure two-photon interference.
A time-tagging device records detection events for two single-photon avalanche photodetectors (SPADs), A and B.
Each building contains a Hanbury Brown-Twiss~\cite{HanburyBrown1956,Kimble1977} setup (not pictured) for measurement of the second-order intensity autocorrelation functions, $g^{(2)}_{\text{ion}}(\tau)$ and $g^{(2)}_{\text{atom}}(\tau)$, of the light from ion and atomic-ensemble sources, respectively.

\begin{figure*}[ht]
    \centering
    \includegraphics[width=1\textwidth]{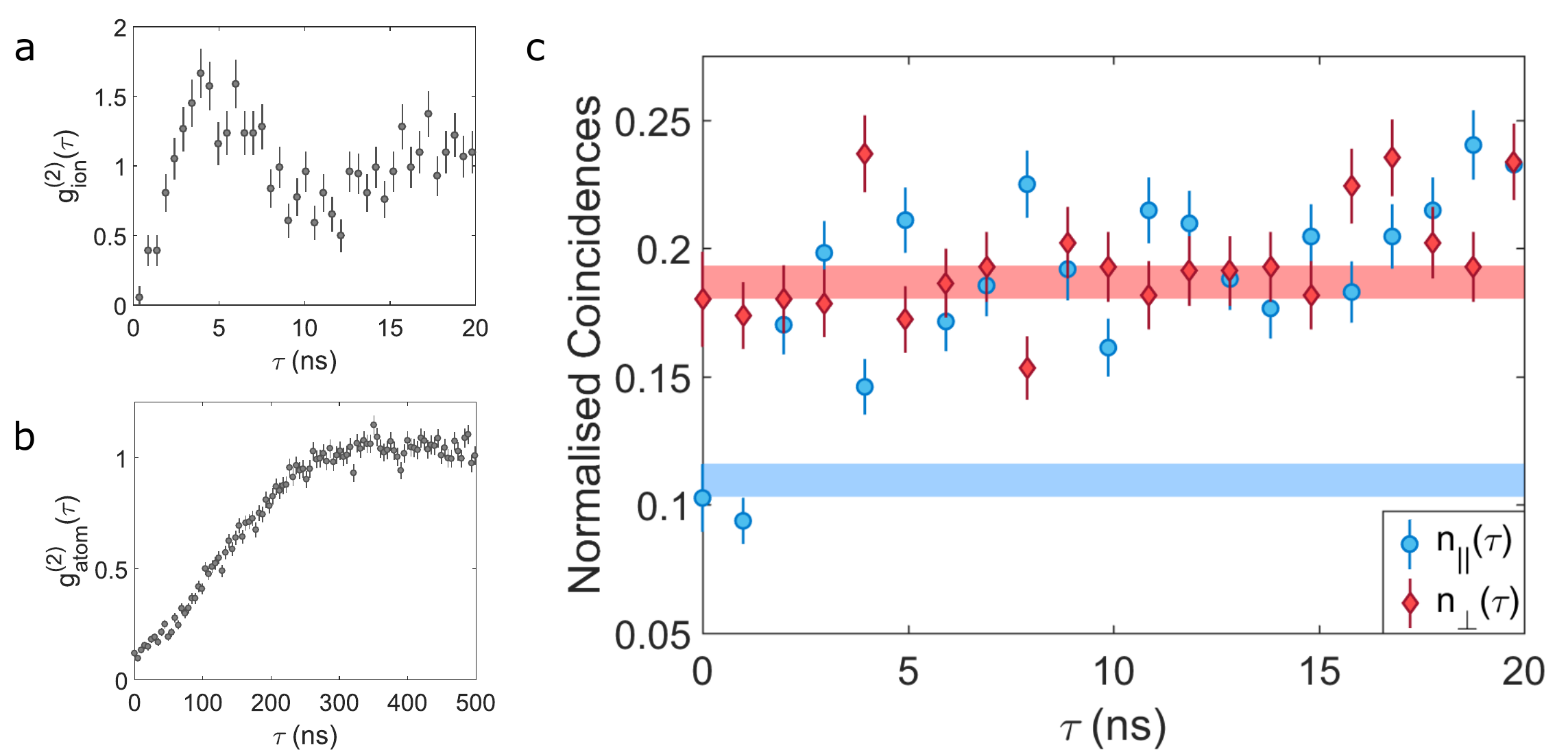}
    \caption{\textbf{Data for stochastic photon production and interference.}
    Second-order intensity autocorrelation functions for the \textbf{a}, ion source, and  \textbf{b}, atomic ensemble source.
    \textbf{c}, Normalised coincidences for the cases where the relative polarisation of the two sources at the interferometer are parallel, $n_\parallel(\tau)$, and perpendicular, $n_\perp(\tau)$, using $1$~ns bins.
    Lower/blue (upper/red) band corresponds to the expected normalised coincidences when photons from the sources are completely indistinguishable (distinguishable).
    For the lower/blue band expected coincidences are entirely due to atomic-ensemble source multiphoton events, while for the upper/red band there is an additional contribution from photon distinguishability.
    Bands indicate the $\pm1\sigma$ confidence interval in this value due to the uncertainty in $g^{(2)}_{\text{atom}}(0)$.
    Data shown accumulated in $\approx 30$ hours.
    In all cases the error bars denote statistical uncertainties.
    All curves shown include background subtraction.} 
    \label{fig:cw_g2_hom}
\end{figure*}

The ion emits single photons near 493 nm via spontaneous emission from the $6\text{P}_{1/2}$ excited state to the $6\text{S}_{1/2}$ ground state. 
A lens collects these photons ($\approx 4\%$ efficiency), and couples them ($\approx 30 \%$ efficiency) into a single-mode fibre (SMF) connected to DFG-1, described in~\cite{siverns2018neutral}. 
We spatially overlap these photons with a strong 1343-nm pump and couple both into a periodically poled lithium niobate waveguide.
Here, DFG converts the 493-nm photons to 780 nm, whilst preserving their quantum statistics~\cite{Kumar90,siverns2018neutral}.
After filtering out undesired light~\cite{siverns2018slow}, we send the converted photons to the HOM interferometer in Building B via a 150-metre polarisation-maintaining fibre (PMF).
To ensure the centre frequency of the photons produced by DFG-1 matches that of the atomic ensemble, DFG-2 is used in an optical beatnote lock setup which feeds back to the pump laser, where 780-nm light from building B acts as a reference.

The atomic-ensemble source uses Rydberg blockade~\cite{Urban2009} to produce single photons, utilising a typical Rydberg polariton experimental layout~\cite{Gorniaczyk2014,Peyronel2012}.
The ground, $\ket{5\text{S}_{1/2},F=2,m_F=2}$, and Rydberg states, $\ket{n\text{S}_{1/2},J=1/2,m_J=1/2}$ are coupled using a two-photon transition, via an intermediate state, $\ket{5\text{P}_{3/2},F=3,m_F=3}$, shown in Fig.~\ref{fig:layout}c. 
The 780-nm probe light that has passed through the cloud is collected and coupled ($\approx70\%$ efficiency) into a PMF.
We operate with Rydberg levels with principal quantum numbers, $n\geq120$, 
where the blockade radius is significantly larger than the probe beam waist, making the medium effectively one dimensional~\cite{Peyronel2012}.
The atomic ensemble has a lifetime of $\approx 1$ s, limited by the background vapour pressure.
Thus, to maintain reasonable atom numbers over the course of the measurements, we periodically reload the ensemble.

First we consider the case where each source continuously produces single photons with stochastic arrival times.
To produce these photons from the ion, we continuously Doppler cool on the $6\text{S}_{1/2}$ - $6\text{P}_{1/2}$ transition, repumping with 650-nm light, see Fig.~\ref{fig:layout}b.
The average count rate of converted photons throughout the experiment, $R_{\text{ion}}$, measured as the sum of counts on SPAD A and B in Building B, is $\approx400$ s$^{-1}$.
Fig.~\ref{fig:cw_g2_hom}a shows $g^{(2)}_{\text{ion}}(\tau)$ for the  resulting 780-nm photon stream.
We measure $g^{(2)}_{\text{ion}}(0)=0.05(8)$ after background subtraction.
Here, the $g^{(2)}_{\text{ion}}$ dip width is set by the effective Rabi frequency ($\approx2\pi \times 100$~MHz) of the driving 493-nm light, which additionally dictates the emitted photon's bandwidth.

\begin{figure*}[ht]
    \centering
    \includegraphics[width=0.8\textwidth]{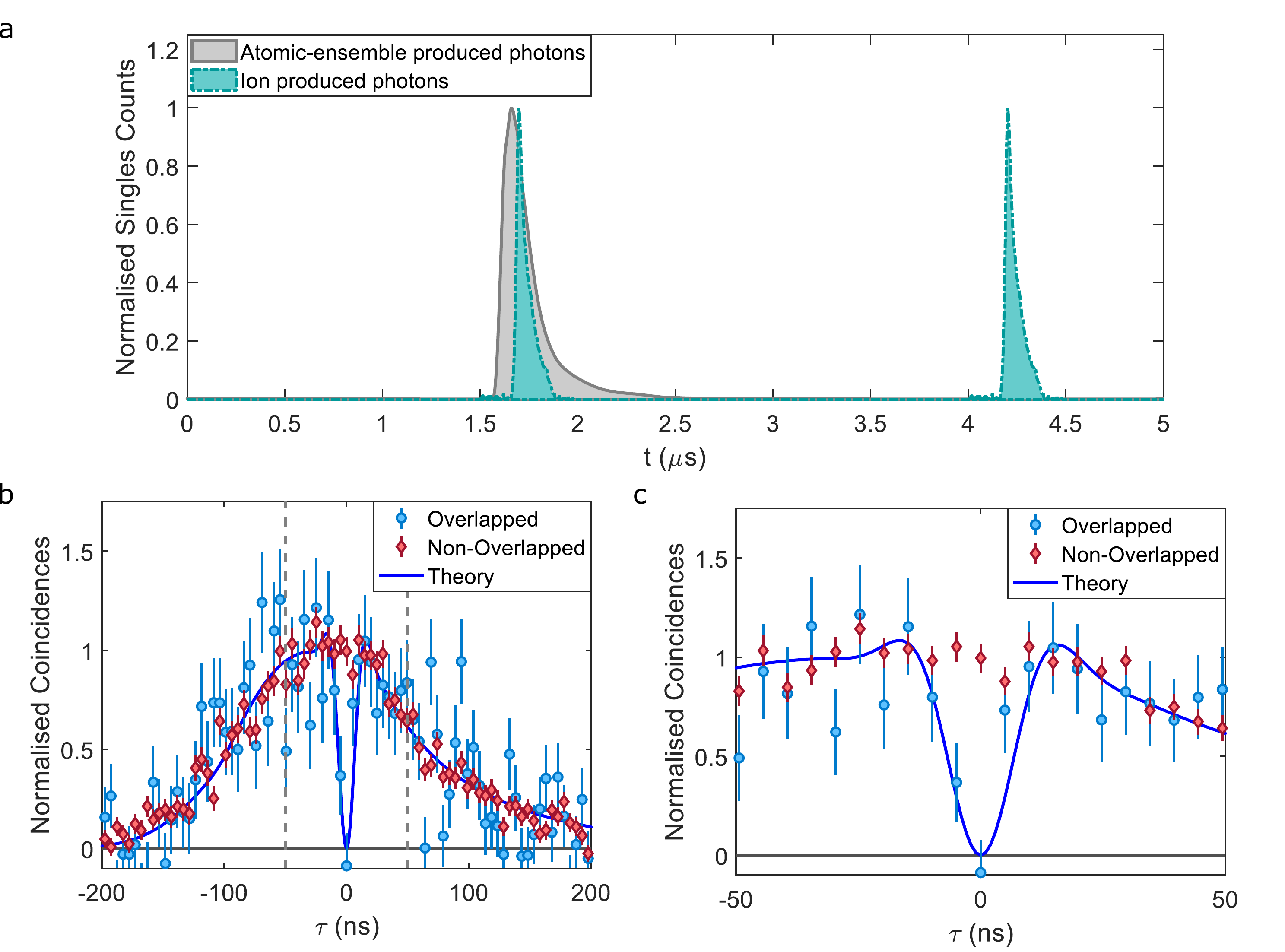}
        \caption{\textbf{On-demand pulse sequence and interference.}
        \textbf{a}, Schematic of pulse sequence for one period. 
        The atomic-ensemble produced photon profile, and ion-produced photon profile at $t\approx4.25$ $\mu s$, are measured directly.
        The ion-produced photon profile at $t\approx1.75$ $\mu s$ is a time shifted copy of that at $t\approx4.25$ $\mu s$ to allow for easy comparison of the photon temporal shapes from the two sources.
        To lessen the effects of small drifts in the relative arrival time of the photons we offset the ion and atomic-ensemble produced photon average arrival (see methods).
        \textbf{b}, and \textbf{c}, Normalised coincidences when the photons from the two sources are temporally overlapped (non-overlapped) shown in blue (red).
        Both curves represent the data after software gating, background subtraction and using $5$~ns bins (see supplementary).
        Dashed lines in \textbf{b} indicate the range shown in \textbf{c}.
        Theory curve obtained taking into account the non-transform limited nature, probabilistic spectrum of the ion-produced photon, and plausible estimates of the relative drift ($2\pi\times 10$~MHz) and offset ($2\pi\times20$~MHz) between the centre frequencies of the photons from the two sources (see supplementary).
        Data presented accumulated over $\approx 22$ hours.
        In all cases the error bars denote statistical uncertainties.}
    \label{fig:od_g2_hom}
\end{figure*}

To produce a stochastic photon stream from the atomic ensemble source, we tune the probe and 479-nm control fields to their respective atomic resonances, $\Delta=\delta=0$, see Fig.~\ref{fig:layout}c.
Rydberg electromagnetically induced transparency (REIT) ensures that only single photons propagate through the medium without large losses~\cite{Peyronel2012}.
In steady-state operation at a high Rydberg level, $n=120$, and large optical depth, $\text{OD}\approx30$, we observe a background subtracted $g^{(2)}_{\text{atom}}(0)=0.119(7)$, shown in Fig.~\ref{fig:cw_g2_hom}b.
We attribute the non-zero value of $g^{(2)}_{\text{atom}}(0)$ to finite probe beam size and input photon flux effects~\cite{Bienias2018, Peyronel2012}.
The width of the $g^{(2)}_{\text{atom}}$ dip is set by the REIT bandwidth~\cite{Peyronel2012}.
However, the majority of the photons exiting the medium have similar spectral bandwidths to the input probe field~\cite{Zeuthen2017}.
We measure an average photon count rate throughout the experiment, $R_{\text{atom}}$, of $\approx 10^4$  $\text{s}^{-1}$, with an experimental duty cycle of 0.56, where the off time is used for reloading.

The background-subtracted normalised coincidences for the HOM interference are shown in Fig.~\ref{fig:cw_g2_hom}c for the cases where the relative polarisation at the interferometer of the photons from the two sources are parallel, $n_{\parallel}(\tau)$,  and  perpendicular, $n_{\perp}(\tau)$.
We define the visibility of the interference:
\begin{equation}
     \label{measured}
    V=\frac{n_{\perp}(0)-n_{\parallel}(0)}{n_{\perp}(0)}
\end{equation}

\noindent
and observe $V = 0.43(9)$ using $1$-ns bins.
For a perfect 50:50 beamsplitter two factors can contribute to a non-unity visibility: multiphoton events from either of the sources, quantified by $g^{(2)}(0)$, and distinguishability.
Multiphoton events decrease the visibility by a factor
\begin{equation}
        \label{expected}
        f_{\text{mp}} =\left[1+\frac{r g^{(2)}_{\text{atom}}(0) + r^{-1} g^{(2)}_{\text{ion}}(0)}{2}\right]^{-1},
\end{equation}
where $r=R_{\text{atom}}/R_{{\text{ion}}}$. 
Equation (\ref{expected}) holds for the case where the photon flux is constant over the experiment (see supplementary), which is a valid approximation for this data.
Given the independently measured $g^{(2)}(0)$ for the sources and ratio, $r$, we determine $f_{\text{mp}} = 0.41(1)$.
The observed $0.43(9)$ visibility can thus be attributed entirely to multiphoton events, and therefore is consistent with perfect bunching of photons from the two sources.
Additionally, we note that $n_{\parallel}(0)$  and  $n_{\perp}(0)$ are in agreement with the values expected from the measured $g^{(2)}(0)$'s, shown in Fig.~\ref{fig:cw_g2_hom}c (see supplementary).
Due to the disparity in the spectral widths of the photons produced by the sources, the width of the HOM dip, seen in Fig.~\ref{fig:cw_g2_hom}c, is almost entirely determined by the temporally narrower ion-produced photon.

To be useful for quantum networking, the photons should arrive on demand in well-defined temporal modes \cite{aspelmeyer2003satellite}.
To this end, we investigate two-photon interference in the case where a single photon from each source arrives at a known time relative to an experimental trigger shared between the two buildings.  

To produce on-demand single photons from the ion, we first prepare it in the $5D_{3/2}$ manifold via optical pumping using 493-nm light.
A pulse of 650-nm light then excites to the $6P_{1/2}$ manifold, from which decay to the $6S_{1/2}$ ground manifold produces a single 493-nm photon~\cite{siverns17}, with measured $g^{(2)}_{\text{ion}}(0)=0(1)\times10^{-2}$  after background subtraction.
We detect a photon at the output of the HOM interferometer with a probability of $\approx2\times10^{-5}$ per attempt.
Photons are emitted from the ion with a nearly exponential decaying temporal profile, with a decay constant ($\approx50$~ns) set by the effective Rabi frequency of the $650$-nm retrieval pulse.
Due to the magnetic bias field ($\approx5$~G) splitting the Zeeman states in the $6S_{1/2}$ and $5D_{3/2}$ levels, combined with the near-equal population distribution in the $5D_{3/2}$ manifold following pumping, the average photon spectrum consists of several  peaks (see supplementary).

For the atomic-ensemble source, we generate on-demand photons using a write and retrieve protocol, similar to that in~\cite{Dudin2012}.
A Rydberg spin wave is written to the cloud using coherent control and probe pulses, detuned far from intermediate resonance ($\Delta \gg \Gamma$, the linewidth of the intermediate state) and close to two-photon resonance ($\delta = 0$).
Rydberg blockade during the write process ensures that a single Rydberg spin wave excitation is stored in the medium.
The control field is tuned close to resonance and then turned on, retrieving the spin wave as a single photon with a spatial mode similar to the input probe light. 
After accounting for background coincidences, we measure $g^{(2)}_{\text{atom}}(0)=0(1) \times 10^{-4}$, with a per-attempt detection probability $\approx3 \times 10^{-2}$ at the outputs of the HOM interferometer.
The temporal profile of the retrieved photon is determined by the control Rabi frequency ($\approx2\pi\times 7$~MHz), intermediate state detuning ($\approx 2\pi\times 7$~MHz), and optical depth ($\approx10$) of the cloud~\cite{Gorshkov2007}.
Fig.~\ref{fig:od_g2_hom}a shows the temporal profile of the atomic-ensemble produced photon, well approximated by a decaying exponential, with a decay constant $\approx120$~ns.

To measure the visibility in a single experimental run, instead of using polarisation to make the photons distinguishable, we use a procedure where the ion-produced photons alternately arrive simultaneously on the beamsplitter with the atomic-ensemble produced photons (with identical polarisation), interleaved with pulses when their arrival times are not overlapped, depicted in Fig.~\ref{fig:od_g2_hom}a.
We use coincidences across several shifted arrival times to correspond to our orthogonal mode reference, to improve statistical noise (see supplementary).
Our experimental sequence consists of requesting photons from the atomic ensemble at a rate of 200~kHz, while the ion produces photons at 400 kHz, triggered via an optical link between the buildings.
We offset the average arrival times of the ion and atomic-ensemble produced photons to mitigate the effects of the small drifts in the relative arrival time of the two sources (see methods).
We operate at an experimental duty cycle of 0.6, with the non data-taking time required to reload the atomic ensemble.

To mitigate noise effects we software gate SPAD A  around a time window containing $\approx80\%$ of the ion-produced photon temporal profile (see supplementary). 
With this gating, we count the coincidences in detection events between SPAD A and B. 
Fig.~\ref{fig:od_g2_hom}b and c show the resulting data for 5 ns bins after subtraction of background coincidences and software gating (see supplementary).
Using equation (\ref{measured}), where $n_{\parallel}$ and $n_{\perp}$ correspond to the the temporally overlapped and non-overlapped coincidences respectively, we calculate a visibility of $1.1(2)$, indicating perfect two-photon interference.
The observed width of the interference dip is narrower than expected when only considering the temporal profile of the photons~\cite{legero2003}.
However, accounting for the multiple peaks in the ion-produced photon spectra, reasonable laser-frequency drifts and average centre-frequency differences of the two photons, we obtain agreement between theory and experiment (see supplementary), seen in Fig.~\ref{fig:od_g2_hom}b and c.

Having observed interference between photons generated from two fundamentally different quantum sources, we now examine our results in the context of hybrid quantum networking.
We consider the entanglement generation scheme in~\cite{simon2003robustentanglement,hucul2015modular}, as a natural extension of our set-up to create a Bell-state analyser, enabling the heralded generation of maximally entangled matter qubits.
With this scheme, the resulting state fidelity, assuming perfect photon-matter entanglement and polarisation discrimination, can be related to the visibility of the two-photon interference $F=(1+V)/2$ (see supplementary).
For the 5-ns bins in Fig.~\ref{fig:od_g2_hom} c, we project $F\approx1$. 
With the measured $\approx40$ bunching events and $\approx21$~hours experimental run time, we infer an entanglement rate of $\approx2$ hour$^{-1}$.
Compromising by using a larger bin we can increase the entanglement rate while decreasing the fidelity \cite{Olmschenk}. 
For example, with a 10-ns bins we estimate an entanglement rate of $\approx6$ hour$^{-1}$ with $F\approx0.9$, still well above the classical limit.
Such entanglement rates are comparable with the first experiments using similar schemes with homogeneous matter qubits~\cite{MonroeRemoteEntanglement,matsukevich2008bell}. 
Additionally, we note that with reasonable improvements to photon collection and detection, entanglement generation rates on the order of several events per minute, with $F>0.9$ is achievable (see supplemental).
The realisation of these projections makes the construction of a hybrid ion-atomic ensemble quantum network attainable. 

%----------------------------------------------------------------------------

%----------------------------------------------------------------------------

\section*{Acknowledgements}

A.C, D.O.-H, A.J.H., S.L.R. and J.V.P. acknowledge support from the United State Army Research Lab's Center for Distributed Quantum Information (CDQI) at the University of Maryland and the NSF Physics Frontier Center at the Joint Quantum Institute.
%is there enough overlap that we don't need to separate names here?
%Do you guys also get funding through PFC?
%asking Qudsia
J.H, J.S. and Q.Q acknowledge support from the United States Army Research Lab's Center for Distributed Quantum Information (CDQI) and the United States Army Research Lab.

\section*{Author Contributions}

A.C., D.O.-H. and A.H. constructed the atomic-ensemble and two-photon interferometer apparatus, and performed atomic-ensemble specific modelling and analysis.
J.H. and J.S. constructed the trapped ion frequency conversion apparatus, and performed ion-specific modelling and analysis.
A.C., J.H., D.O.-H., J.S. and A.H. acquired the data.
A.C. performed the two-photon interference data analysis.
J.H. performed entanglement rate and fidelity calculations.
E.G. helped with two-photon experimental techniques and interpretation of data.
J.P., Q.Q. and S.R. supervised the project;
All authors contributed to discussions about the experiment and to the writing of the manuscript.

\section*{Additional Information}
The identification of commercial products in this paper does not imply recommendation or endorsement by the National Institute of Standards and Technology or the Army Research Laboratory, nor does it imply that the
items identified are necessarily the best available for the purpose.

\section*{Methods}

\subsection*{Trapped Ion}
We confine and Doppler cool a single $^{138}$Ba$^+$ ion using a radio-frequency Paul trap \cite{siverns2017traps} and 493-nm light.
An additional laser at 650-nm is used as a re-pumper.
In both modes of photon production, the ion emits single photons at 493-nm, with any photons produced at 650-nm filtered out by the frequency conversion setup.
For the case of stochastic photon generation, the ion is constantly Doppler cooled with Rabi frequencies of $\approx2 \pi \times 25$ MHz and $\approx2 \pi \times11$ MHz for the 493-nm and 650-nm beams respectively.
The detunings of these beams are $\approx 2\pi \times (-99)$ MHz and $\approx 2\pi \times 29$ MHz respectively.
In the case of on-demand single-photon generation, we use a process similar to that described in \cite{siverns2018slow}. 
First, the ion is pumped equally into each Zeeman sub-level of the $D_{5/2}$ manifold using only the 493-nm laser for 750 ns.
This light is then turned off, and we wait for for 60 ns to ensure any laser scatter is not detected during the photon extraction phase.
A 200 ns pulse of 650-nm light, separate from that used during Doppler cooling, is then used to excite the ion to the $P_{1/2}$ manifold, from which a 493-nm photon may be emitted ($\approx75$\% branching ratio). 
This light is linearly polarised, propagates along the quantisation axis, with a Rabi Frequency of $\approx 2\pi \times$ 30 MHz and detuning of $\approx 2\pi \times 29$ MHz. 
The 650-nm light is turned off, and after a short period (60 ns) with no light, the ion is Doppler cooled (for a minimum of 500 ns), until a trigger is received from the atomic ensemble lab, and the process is repeated.

Photon collection is performed with a custom coated, ex-vacuo 0.4 NA objective, corresponding to about 4\% collection of light emitted by the ion.
The light is then coupled into a SMF ($\approx35\%$ coupling efficiency) which is connected to the DFG-1 setup.
The total photon collection efficiency out of the SMF is $\approx1\%$.
The collected  photons are spatially combined with the pump laser on a dichroic mirror before passing through a 20X objective and coupled into a wave guide in the periodically poled lithium niobate crystal, DFG-1 (SRICO Model: 2000-1005). 
The converted and unconverted photons, as well as pump light, exit through a fibre butt-coupled and glued to the output of the wave guide.
We pass the light from DFG-1 through a set of interference filters (two each of Semrock: LL01-780-25 and FF01-1326/SP-25) and a volume Bragg grating (OptiGrate BP-785) to filter out the pump, noise, and unconverted photons.
Finally, the remaining 780-nm single photons are passed through a polarisation filter before being coupled into the PMF connecting to Building B. 
The single-photon conversion efficiency, measured as the ratio between the number of output 780-nm photons after the Bragg grating and the number of 493-nm photons before combination with the pump, is $\approx$ 10\% on average, with fluctuations that we attribute to photo refractive effects caused by the high intensity pump light~\cite{fejerphotorefractive,kashinphotorefractive,fujiwara1992photorefractive}.
Due to these effects, the experimental run time is limited to about 10 hr before the pump must be turned off for an extended period of time (10 hr) to allow the crystal to recover.

\subsection*{Atomic ensemble}

We trap and cool $^{87}$Rb atoms in a magneto-optical trap (MOT).
After a period of grey optical molasses~\cite{Boiron1995}, cooling the ensemble to $\approx 10$ $\mu$K, we load the atoms into a crossed optical dipole trap, consisting of three 1004-nm beams, two that cross at a shallow angle and a third ``dimple" beam that is nearly perpendicular, with all beams in the same plane. 
By varying loading time of the MOT and relative power in the shallow cross and dimple beams we can adjust the optical depth up to $\approx30$, and the rms radius of the atomic cloud, parallel to the direction of the probe, from $25$ to $45$ $\mu $m.
After loading into the dipole trap we optically pump the atoms into the $\ket{5\text{S}_{1/2},F=2,m_F=2}$ state, using $\sigma_+$ polarised light addressing the $F=2$ to $F'=2$ D1 transition.

For both methods of photon generation we couple the ground and Rydberg state using a 780-nm probe field, which couples the $\ket{5\text{S}_{1/2},F=2,m_F=2}$ ground state to the $\ket{5\text{P}_{3/2},F=3,m_F=3}$ intermediate state, and a \mbox{479-nm} control field, which couples the $\ket{5\text{P}_{3/2},F=3,m_F=3}$ intermediate state to the $\ket{n\text{S}_{1/2},J=1/2,m_J=1/2}$ Rydberg state.
The lasers that generate the probe and control light are both frequency stabilised using an ultra-low expansion (ULE) cavity (linewidth $<10$ kHz).
We use the probe light exiting the ULE cavity to reduce phase noise \cite{Leseleuc2018}.

The probe beam is focused down to a $1/e^2$ waist of $\approx3.3$~$\mu$m, with the focus located at the centre of the optical dipole trap.
The control beam is counter-propagated relative to the probe and focused down to a beam waist of $\approx20$~$\mu$m.
After the cloud, we pass the probe light through a set of 780-nm centre frequency bandpass filters (Alluxa 780-1 OD6 and Semrock Brightline 780/12) prior to coupling into the PMF.

As discussed in the main text, a continuous, stochastic stream of photons is produced by passing a coherent 780-nm probe beam through the atomic-ensemble under REIT conditions.
For optimal single photon filtering~\cite{Peyronel2012} we use the maximum optical depth, OD~$\approx30$, for which the cloud has an rms radius of $\approx40$~$\mu$m along the probe direction.
We use a Rydberg principal quantum number $n$ = 120, for which we obtain a control Rabi frequency $\approx 2\pi\times 11$~MHz, blockade radius $\approx20$~$\mu$m, defined as in~\cite{Bienias}, and on-resonance REIT transmission $\approx50$\%.

For the on-demand protocol we use a Rydberg principal quantum number, $n=139$.
To write the spin wave we pulse the control, Rabi frequency $\Omega_c\approx 2\pi\times 7$~MHz, and probe field, Rabi frequency $\Omega_p\approx 2\pi\times 1$~MHz, on for $\approx370$~ns.
The write is performed near two-photon resonance but detuned from intermediate state resonance, $\Delta\approx2\pi\times 50$~MHz.
An additional $20$-fold enhancement to the two-photon Rabi frequency, $\Omega_{\text{2-photon}}\approx \Omega_c\Omega_p/(2\Delta)$, is observed due to Rydberg blockade~\cite{Saffman2010}.
To ensure that only a single spin wave is written we work with a longitudinally short cloud with an rms radius $\approx30$~$\mu$m, much smaller than the blockade radius, $\approx60$~$\mu$m, defined in terms of the van der Waals $C_6$ coefficient and two-photon Rabi frequency as $\left(C_6/\Omega_{\text{2-photon}}\right)^{1/6}$.
For this length cloud the optical depth is $\approx13$.
After the write process, the spin wave is held for $\approx350$~ns to allow time to switch the control field frequency, which is controlled by a double-passed acousto-optic modulator (AOM), to $\Delta\approx2\pi\times 7$~MHz from resonance.
The control field is then turned on and the spin-wave is retrieved as a photon.
We use an AOM as a shutter after the cloud to ensure that the photons from the probe write pulse to do not saturate the SPADs.

\subsection*{Frequency Lock}
To ensure the converted ion-produced photons are at a similar frequency as those produced by the atomic ensemble, we use a second frequency conversion setup, DFG-2. 
Laser light at 493-nm, with a known detuning from the photons emitted by the ion ($\pm$ 10 MHz), is combined with the same pump light used in DFG-1, producing continuous wave light at 780-nm.
The output frequencies of DFG-1 and DFG-2 can both be controlled by changing the pump frequency.
The 780-nm light from DFG-2 is combined with frequency-locked 780-nm light from building B onto a fast photodetector (Electro-Optics Technology ET-2030A), with which we measure an optical beat note.
By using the beat note to produce an error signal used to feed back to the pump laser's frequency control, we stabilise and set the frequency of the output 780-nm light from both DFG setups.
Uncertainties in the centre frequency of the converted 780-nm single photons were present in the experiment due to uncertainties in the ion spectroscopy, and drifts in the 493-nm and 650-nm laser wavemeter locks.
These two factors affect the two-photon interference and are investigated in the supplementary material.

\subsection*{HOM interferometer}

Light from both sources is transmitted to the interferometer setup by PMF.
At the output of each fibre we use a polarising beamsplitter (PBS) to clean the polarisation of the light before it passes to the 50:50 beasmplitter.
For the atomic-ensemble source a HWP after the PBS allows us to adjust the relative polarisation of the two sources at the 50:50 beamsplitter.
We couple the output ports of the 50:50 beamsplitter to a pair of SMFs with similar mode field diameters to the input PMFs, which are connected to a pair of SPADs (Excelitas SPCM-780-13).
Immediately prior to both output SMFs we use a bandpass filter (Semrock Brightline 780/12) to remove stray light.
We use a time-tagger (Roithner-Laser TTM8000) to record timestamps for SPAD detection events, from which we use software to calculate coincidences.

\subsection*{On-Demand Synchronisation}

We operate in master-slave configuration with the atomic-ensemble lab, in building B, as the master and the ion lab, in building A, as the slave.
In the ensemble lab we generate 1064-nm optical pulses using an AOM with laser light.
These are sent over fibre to the ion lab where the optical pulse is converted to TTL, using a high bandwidth (ThorLabs PDA05CF2) photodiode, which triggers photon production.
Due to drifts in the power of the 1064-nm optical pulse, we observe small drifts ($\leq 20$~ns  over several hours) in the ion-produced photon arrival time relative to that generated by the atomic-ensemble.
To ensure the photon profiles overlap, even with these drifts, we offset the average arrival time of the ion produced photon $\approx+40$ ns relative to atomic-ensemble produced photon.
Calculations indicate that with this offset such temporal drifts have negligible effect on the two-photon interference.
We observed no measurable drift between the temporally overlapped and non-overlapped photons produced by the ion, with a temporal separation of $2.5$~$\mu$s.

Along with events on SPAD A and B we additionally record timestamps for an electronic reference, which defines an absolute time reference within the $5$-$\mu$m pulse cycle, shown in Fig.~3a.
This reference was provided by the same electronics that controlled the arrival time of the photons produced by the two systems.
Throughout the experiment we observed no drift between the arrival time of the atomic-ensemble produced photon and the electronic reference.

\newpage

\onecolumn

\renewcommand{\thesubsection}{S.\arabic{subsection}}
\renewcommand{\theequation}{S\arabic{equation}}
\renewcommand{\thefigure}{S\arabic{figure}}
\renewcommand{\thetable}{S\arabic{table}}
\setcounter{equation}{0}
\setcounter{figure}{0}
\section*{Supplementary}

\subsection{Derivation of reduction of visibility due to multiphoton events}\label{sec:visDeriv}

\begin{figure}[h]
    \centering
    \includegraphics[width=0.45\textwidth]{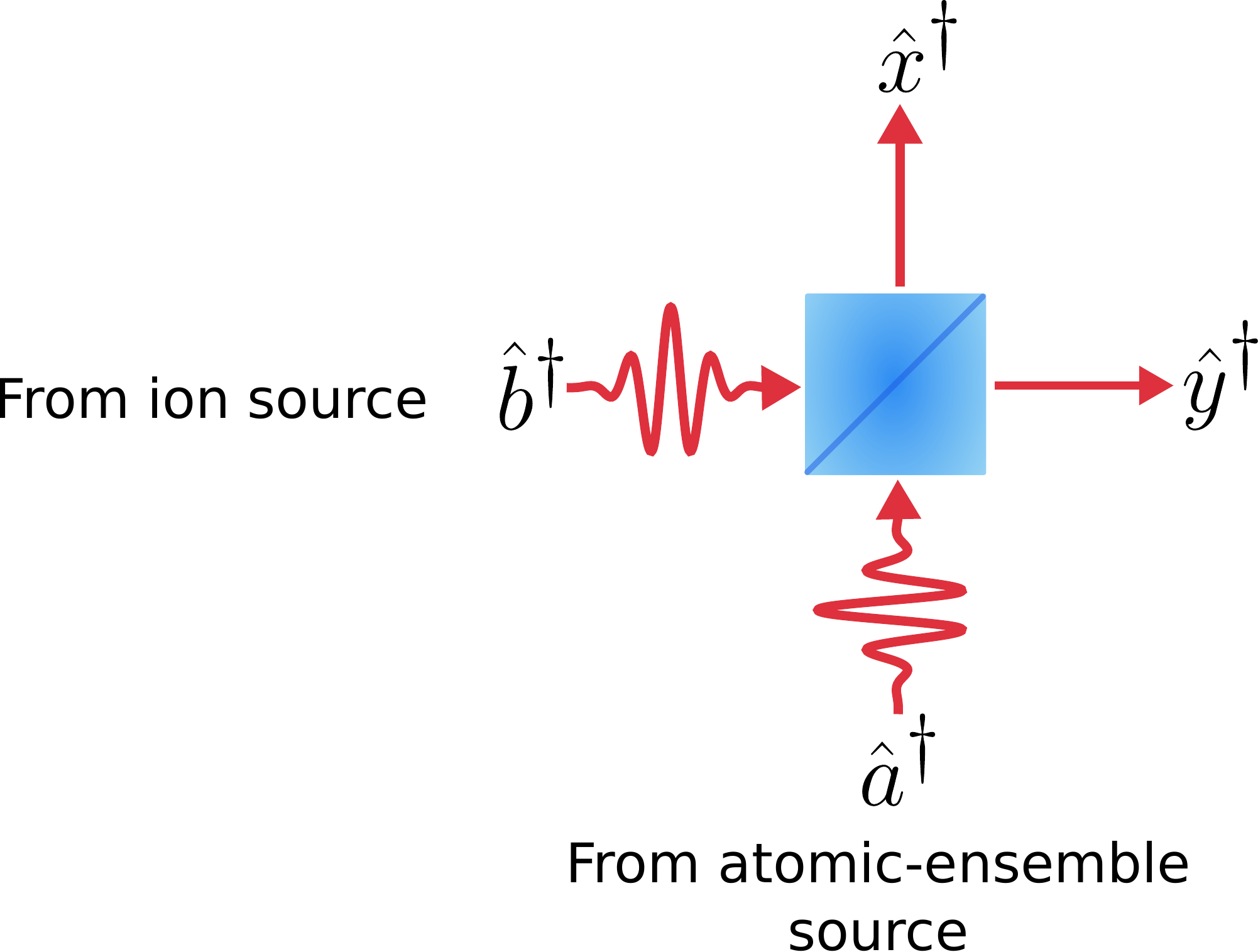}
    \caption{\textbf{Two photons incident on a beamsplitter.} Input photons are represented by the raising operators $\hat{a}^\dagger$ and $\hat{b}^\dagger$ for the atomic-ensemble and ion sources respectively, with $\hat{x}^\dagger$ and $\hat{y}^\dagger$ representing photons exiting each of the two output ports. 
    }
    \label{fig:beamsplitter}
\end{figure}

Here, we derive the expected visibility of the interference between two photons incident on a perfect 50:50 beamsplitter.
The relation between the input and output bosonic operators is given by:

\begin{equation}
  \begin{pmatrix}
    \hat{a}^\dagger\\ \hat{b}^\dagger
    \end{pmatrix}
    \rightarrow \frac{1}{\sqrt{2}}\begin{pmatrix}
    1 & i\\ i & 1
    \end{pmatrix}
     \begin{pmatrix}
    \hat{x}^\dagger\\ \hat{y}^\dagger
    \end{pmatrix}.
\end{equation}

Initially, we consider the case where a single photon from the atomic ensemble, represented by $\hat{a}_{\text{atom}}^\dagger$, and a single photon from the ion, represented by $\hat{b}_{\text{ion}}$, are present at separate inputs of beamsplitter.
For such an input state we have:
\begin{equation}
    \begin{split}
        \ket{1_{\text{atom}},1_{\text{ion}}}_{\text{in}} &= \hat{a}_{\text{atom}}^\dagger\hat{b}_{\text{ion}}^\dagger\ket{0,0}_{\text{in}}\\
        &\rightarrow \frac{1}{2}\left( \hat{x}_{\text{atom}}^\dagger + i\hat{y}_{\text{atom}}^\dagger \right)\left( i \hat{x}_{\text{ion}}^\dagger +  \hat{y}_{\text{ion}}^\dagger \right)\ket{0,0}_{\text{out}}.
    \end{split}
    \label{photon beamsplitter}
\end{equation}
\noindent
To take into account the distinguishability between the two input photons, we define:

\begin{equation}
    \begin{split}
        \hat{x}_{\text{atom}}^\dagger &= \hat{x}^\dagger\\
        \hat{y}_{\text{atom}}^\dagger &= \hat{y}^\dagger\\
        \hat{x}_{\text{ion}}^\dagger &= \sqrt{c}\ \hat{x}^\dagger+\sqrt{1-c}\ \hat{x}_n^\dagger\\
        \hat{y}_{\text{ion}}^\dagger &= \sqrt{c}\ \hat{y}^\dagger+\sqrt{1-c}\ \hat{y}_n^\dagger,
    \end{split}
    \label{redefine}
\end{equation}

\noindent
where $c$ is a real number, $0\leq c \leq 1$, and parameterises the mode overlap of the two photons.
Here $\hat{x}_n^\dagger$ and $\hat{y}_n^\dagger$ consist of all modes orthogonal to $\hat{x}^\dagger$ and $\hat{y}^\dagger$, respectively, i.e. $0=\bra{0}\hat{x}_n^\dagger\hat{x}^\dagger\ket{0}=\bra{0}\hat{y}_n^\dagger\hat{y}^\dagger\ket{0}$.
From equation (\ref{photon beamsplitter}) and (\ref{redefine}) we find:

\begin{equation}
    \begin{split}
        \ket{1_{\text{atom}},1_{\text{ion}}}_{\text{in}} \rightarrow\ket{\psi_{\text{out}}}&= \frac{i}{2}\left(\sqrt{2c}\ket{2,0}_{\text{out}}+\sqrt{1-c}\ket{1\&1_n,0}_{\text{out}} + \sqrt{2c}\ket{0,2}_{\text{out}}+\sqrt{1-c}\ket{0,1\&1_n}_{\text{out}}\right)\\
        &+\frac{\sqrt{1-c}}{2}\left(\ket{1,1_n}_{\text{out}}-\ket{1_n,1}_{\text{out}}\right),
    \end{split}
\end{equation}

\noindent
where we use the notation $\ket{1\&1_n,0}_{\text{out}}$ and $\ket{0,1\&1_n}_{\text{out}}$ to denote instances where the photons exit the same port of the beamsplitter, but are otherwise in orthogonal modes. 
The probability, $P\left({\ket{1,1}_{\text{in}}\rightarrow 1,1}\right)$, of finding a photon at both output ports is then given by:

\begin{equation}
    \begin{split}
        P_{{\ket{1,1}_{\text{in}}\rightarrow 1,1}} &= \abs{\braket{1,1}{\psi_{\text{out}}}}^2 + \abs{\braket{1,1_n}{\psi_{\text{out}}}}^2 + \abs{\braket{1_n,1}{\psi_{\text{out}}}}^2\\
        &= \frac{\left(1-c\right)}{2}.
    \end{split}
\end{equation}

From this, we can see that a value of $c=1$ corresponds to perfect interference, with zero probability of finding photons at both output ports simultaneously.
Similarly, a value of $c=0$ corresponds to no interference between the photons, with equal probability for the photons to exit the same port, or separate ports.

We now repeat the same procedure for the situation where two photons from the atomic-ensemble source are present at the beamsplitter with none from the ion source, i.e. $\ket{2_{\text{atom}},0}_{\text{in}}$:

\begin{equation}
    \begin{split}
        \ket{2_{\text{atom}},0}_{\text{in}} &= \frac{1}{\sqrt{2}} \left(\hat{a}_{\text{atom}}^\dagger\right)^2\ket{0,0}_{\text{out}}\\
        &\rightarrow \frac{1}{2\sqrt{2}}\left(\hat{x}^\dagger + i\hat{y}^\dagger \right)^2\ket{0,0}_{\text{out}}\\
        &\rightarrow \frac{1}{2}\left(\ket{2,0}_{\text{out}}-\ket{0,2}_{\text{out}}+i\sqrt{2}\ket{1,1}_{\text{out}}\right).
    \end{split}
\end{equation}

\noindent
Using a similar procedure as for the $\ket{1,1}$ input state, the probability of finding a photon at both output ports is $P_{\ket{2,0}_{\text{in}}\rightarrow 1,1} = 1/2$. 
From symmetry this is the same for the case of the $\ket{0,2_{\text{ion}}}_{\text{in}}$ input state, i.e. $P_{\ket{0,2}_{\text{in}}\rightarrow 1,1} = 1/2$.

For near-single photon sources with low single photon detection probability, such as those used in this work, input states with total photon number $> 2$ occur with negligible probability.
We now calculate the coincidence rate, ignoring such terms:

\begin{equation}
    \label{coinc_rate}
    \mathcal{R}(\tau=0) = \frac{P_{\text{atom}}P_{\text{ion}}}{\Delta\tau} P_{\ket{1,1}_{\text{in}}\rightarrow 1,1} + \frac{P_{2\times \text{atom}}}{\Delta\tau} P_{\ket{2,0}_{\text{in}}\rightarrow 1,1} + \frac{P_{2\times \text{ion}}}{\Delta\tau} P_{\ket{0,2}_{\text{in}}\rightarrow 1,1},
\end{equation}

\noindent where $P_{\text{atom}(\text{ion})}$ and $P_{2\times \text{atom}(\text{ion})}$ are the probabilities of having a single and two photons from the specified source in a time interval, $\Delta\tau$, respectively.

For the case of continuously produced photons, we can rewrite the single photon probability in terms of the singles rates, $R_{\text{atom}(\text{ion})}$, $P_{\text{atom}(\text{ion})} = R_{\text{atom}(\text{ion})}\Delta\tau$.
Additionally we can make the approximation $g^{(2)}_{\text{atom}(\text{ion})}(0)\approx 2 P_{2\times \text{atom}(\text{ion})}/P_{\text{atom}(\text{ion})}^2$ which is nearly exact in the limit of small photon flux~\cite{Stevens2013}, as is true for our experiment.
Equation (\ref{coinc_rate}) can then be written:

\begin{equation}
    \begin{split}
        \mathcal{R}(0) &= \Delta\tau R_{\text{atom}}R_{\text{ion}}\left[\frac{(1-c)}{2}+\frac{1}{4}\left(\frac{R_{\text{atom}}}{R_{\text{ion}}} g^{(2)}_{\text{atom}}(0) + \frac{R_{\text{ion}}}{R_{\text{atom}}} g^{(2)}_{\text{ion}}(0)\right)\right]\\
        &= \Delta\tau R_{\text{atom}}R_{\text{ion}}\left[\frac{(1-c)}{2}+\frac{1}{4}\left(r g^{(2)}_{\text{atom}}(0) +r^{-1} g^{(2)}_{\text{ion}}(0)\right)\right],
    \end{split}
    \label{coinc_rate_exp}
\end{equation}

\noindent
where $r=R_{\text{atom}}/R_{{\text{ion}}}$. From equation (\ref{coinc_rate_exp}) we can calculate the expected normalised coincidences as a function of $c$, $g^{(2)}(0)$ for the two sources and $r$:

\begin{equation}
    \begin{split}
        n(0) &= \frac{\mathcal{R}(0)}{\eval{\mathcal{R}(0)}_{c=0,\,g^{(2)}_{\text{atom}}(0)=1,\,g^{(2)}_{\text{ion}}(0)=1}}\\
        &= \frac{2\left(1-c\right) + r g^{(2)}_{\text{atom}}(0) +r^{-1} g^{(2)}_{\text{ion}}(0)}{2+r +r^{-1}}.
    \end{split}
    \label{norm_coinc}
\end{equation}

We use equation (\ref{norm_coinc}) to calculate the bands shown in Fig.~\ref{fig:cw_g2_hom}c of the main text, where we set $c=1$ for the case of perfect interference and $c=0$ in the case of no interference.
Assuming no interference in the perpendicular case, and making no assumptions about the overlap in the parallel case,  we can calculate an expected visibility using equation (\ref{measured}) of the main text:

\begin{equation}
    \begin{split}
        V_{\text{exp}} &= \frac{n_{\perp}(0)-n_{\parallel}(0)}{n_{\perp}(0)}\\
        &= \frac{\eval{n(0)}_{c=0}-n(0)}{\eval{n(0)}_{c=0}}\\
        &= 1-\frac{\left(2(1-c) + r_\parallel g^{(2)}_{\text{atom}}(0) +r_\parallel^{-1} g^{(2)}_{\text{ion}}(0)\right)\left( 2+r_\perp +r_\perp^{-1} \right)}{\left( 2+r_\perp g^{(2)}_{\text{atom}}(0) +r_\perp^{-1} g^{(2)}_{\text{ion}}(0) \right) \left( 2+r_\parallel +r_\parallel^{-1} \right)}.
    \end{split}
    \label{visibility}
\end{equation}

\noindent
If we assume the rates are similar in the parallel and perpendicular measurements, $r=r_\perp=r_\parallel$, equation (\ref{visibility}) reduces to:

\begin{equation}
    V_{\text{exp}} = c\left[1+\frac{r g^{(2)}_{\text{atom}}(0) + r^{-1} g^{(2)}_{\text{ion}}(0)}{2}\right]^{-1}.
    \label{eqn:visibility_coal}
\end{equation}

In the limit that $g^{(2)}_{\text{atom}}(0)=g^{(2)}_{\text{ion}}(0)=0$ we see that the visibility is equal to the mode overlap of the photons from the two sources, $c$.
However, given a finite $g^{(2)}(0)$ for either source the visibility is reduced, due to multi-photon events, by a factor:

\begin{equation}
    f_{mp} = 1+\frac{r g^{(2)}_{\text{atom}}(0) + r^{-1} g^{(2)}_{\text{ion}}(0)}{2}
\end{equation}

\noindent
which is equation (\ref{expected}) of the main text.

\subsection{Calculation of on-demand coincidences}

\begin{figure*}[!h]
    \centering
    \includegraphics[width=1\textwidth]{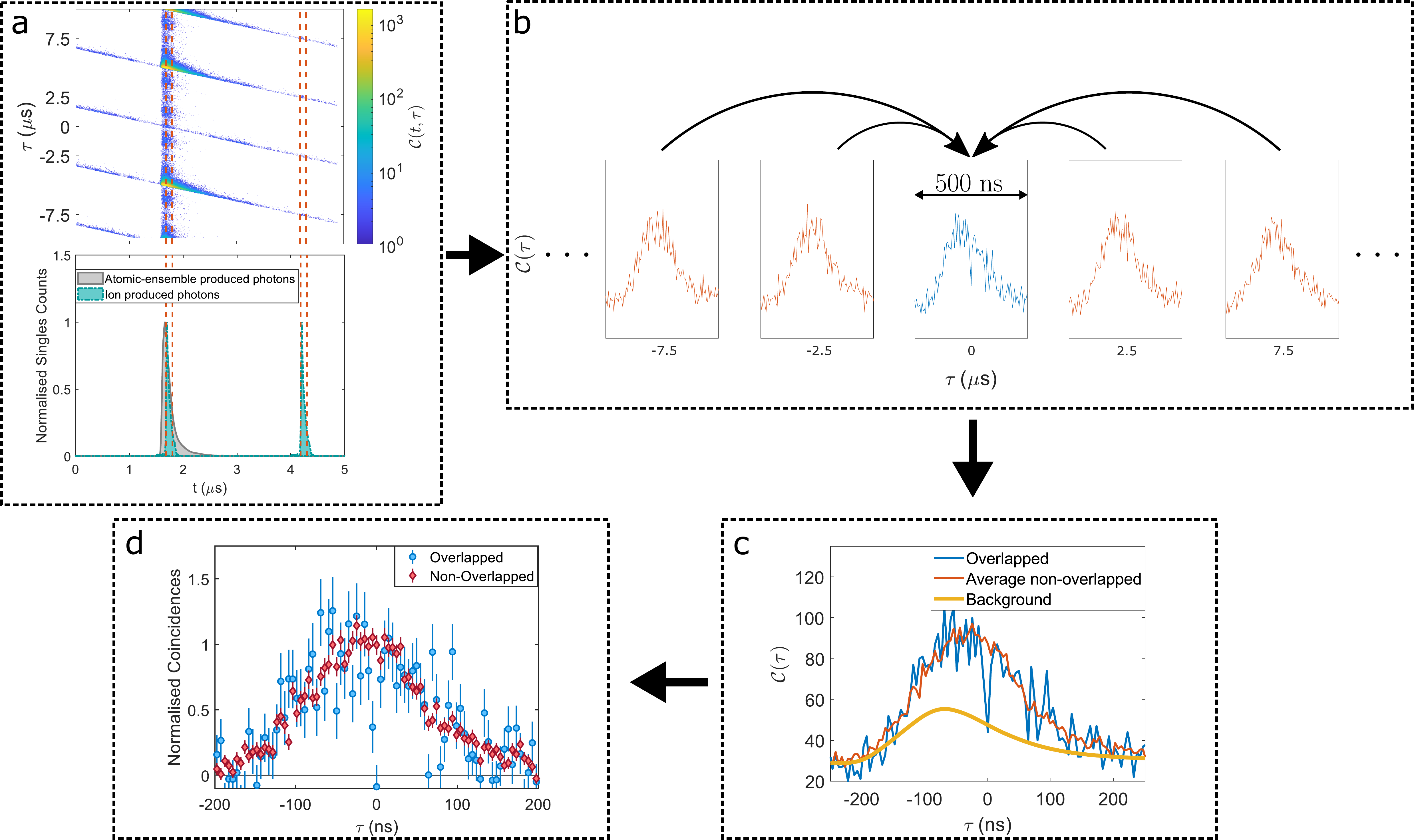}
    \caption{\textbf{Method for calculating coincidences for on-demand produced photons.}
    Starting from \textbf{a}, coincidences are calculated between the two SPADs as a function of relative time between events on the two SPADs, $\tau$, and the absolute time within the pulse sequence, $t$. Additionally we determine the counts for the two sources.
    Here, the atomic-ensemble produced photon shape, and ion-produced photon shape at $t\approx4.25$ $\mu s$, are measured directly, while ion-produced photon profile at $t\approx1.75$ $\mu s$ is a time shifted copy of that at $t\approx4.25$ $\mu s$.
    We select a window, indicated by the dashed-orange lines, around the two ion-produced photons and sum as a function of $t$ to obtain $\mathcal{C}(\tau)$, shown in \textbf{b}.
    We take an ensemble of $\mathcal{C}(\tau)$ curves, temporally shift them by $\pm (2.5 + 5k)$~$\mu$s, with $k$ an integer, and average them together to obtain our non-overlapped curve, shown in \textbf{c}.
    Background coincidence subtraction is then performed and the resulting curves are normalised to produce the final figure shown in \textbf{d}.
    In \textbf{b} through \textbf{d} we draw connecting lines between data points to guide the eye.
    %DISCONNECT THE LINES IN LOWER LEFT.
    } 
    \label{fig:coinc_explanation}
\end{figure*}

\noindent
Due to the relatively low flux of ion-produced photons at the output of the interferometer, we perform software gating to improve the signal-to-noise ratio of the coincidences for the on-demand interference.
To perform this gating we first determine the arrival times, relative to the electronic clock, of the two ion-produced photons, shown in Fig.~\ref{fig:coinc_explanation}a around $\approx 1.7$, $4.2$~$\mu$s.
We calculate coincidences, $\mathcal{C}(t,\tau)$,  between the two SPADs as a function of the relative time $\tau$ between events on SPAD A and B, and the time $t$ between the event on SPAD A and the electronic clock, shown in Fig. \ref{fig:coinc_explanation}a.
We take two windows, denoted by the regions between the dashed-orange lines in Fig~\ref{fig:coinc_explanation}a, and calculate $\mathcal{C}(\tau) = \sum_{t\in t_w}\mathcal{C}(t,\tau)$, where $t_w$ are the set of times in the windows.
For our data we chose a window size that provided a good compromise between data accumulation rate and signal-to-noise, encompassing $\approx80$\% of the area of the ion-produced photon.
We note that this method is equivalent to a physical gate on the SPADs over the time period of integration.

Due to the large disparity ($\approx\times10^3$) in the ion and atomic-ensemble-produced photon detection probabilities, coincidences between any two independent ion produced photons are negligible.
Therefore ignoring background coincidences, which will be discussed later, features in $\mathcal{C}(\tau)$ around $\tau=0$ arise from instances where the atomic-ensemble-produced and ion-produced photons are overlapped, while features around $\tau=\pm(2.5+5k)$~$\mu$s for $k\in \mathbb{Z}$ arise from instances where the two photons are not overlapped.
To directly compare the cases where the photons are overlapped to the case where they are non-overlapped, we temporally  shift the $\mathcal{C}(\tau)$ curve.
Additionally, we average together several of these temporally shifted data in order to reduce our uncertainties for the non-overlapped case. Procedurally, we take a set of $\mathcal{C}(\tau')$ curves and shift each curve by an amount $\tau_k=(2.5+5k)$~$\mu$s to obtain a set $\left\{\mathcal{C}(\tau'+\tau_k) \right\}$.
We then average to obtain the non-overlapped curve, $\left<\mathcal{C}(\tau=\tau'+\tau_k) \right>_k$, shown in Fig.~\ref{fig:coinc_explanation}c.
For the data presented in the main text the non-overlapped curve was constructed using $-5\leq k \leq 4$.

We calculate the expected background coincidence curve, shown in Fig.~\ref{fig:coinc_explanation}c, from the independently measured experimental singles and background rates from the two SPADs.
For this calculation the same gating described above is used.
To obtain the final coincidence curves, shown in Fig. \ref{fig:coinc_explanation}d, the expected background is subtracted from the overlapped and non-overlapped coincidences and the resulting curves  scaled by the same factor.

\subsection{Effects of experimental imperfections on on-demand interference}

Here we calculate coincidence profiles and compare them to the experimental observations for the on-demand produced photons.
For the following we shall consider the case where we have two single photons incident on two different input ports of a 50:50 beamsplitter, as in Fig.~\ref{fig:beamsplitter}.
We assume that the photons have the same transverse-spatial and polarisation mode but may have different spatio-temporal modes, $\zeta_i$. 
The probability of detecting a photon at time $t_0$ in one detector followed by a detection in the other detector at time $t_0 + \tau$ is~\cite{legero2003}:

\begin{equation}
    P(t_0,\tau) = \frac{1}{4} \left|\zeta_{\text{atom}} (t_0)\zeta_{\text{ion}}(t_0+\tau) - \zeta_{\text{atom}}(t_0+\tau)\zeta_{\text{ion}}(t_0)\right|^2.
\end{equation}

\begin{figure*}[h]
    \centering
    \includegraphics[width=0.95\textwidth]{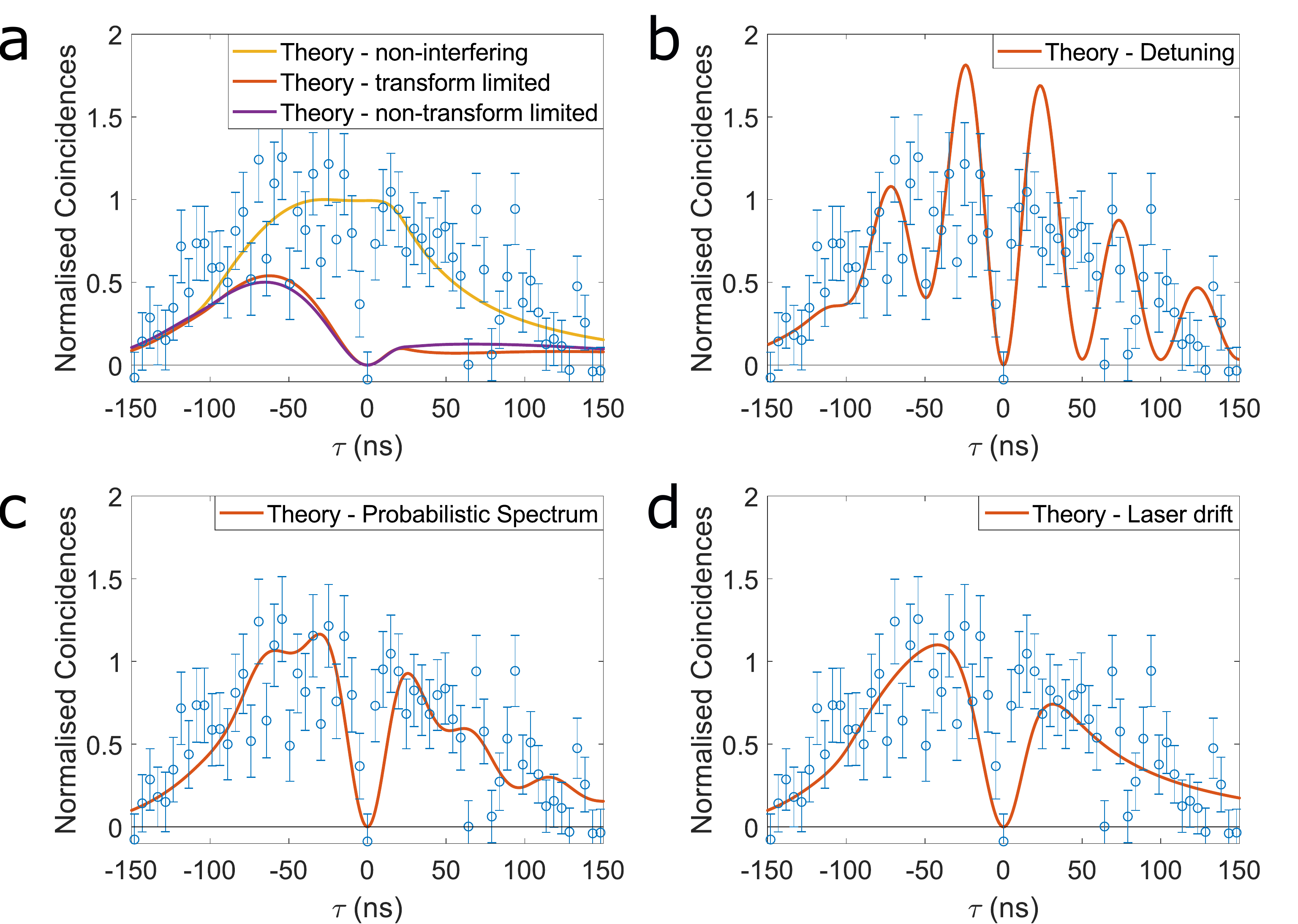}
        \caption{\textbf{Theory accounting for various experimental imperfections.}
        In each panel we show the background-subtracted experimental data along with theory which, in each case, considers a different potential experimental imperfection. The error bars on the experimental data denote statistical uncertainties.
        \textbf{a}, Accounts for the non-transform limited nature of the ion-produced photon due to the branching ratio of the $6P_{1/2}$ state.
        \textbf{b}, constant centre frequency difference ($\Delta \omega = 2\pi\times20$~MHz) between the photons from the two sources.
        \textbf{c}, the probabilistic spectrum of the ion-produced photon due to the Zeeman splitting (see Fig. \ref{fig:BaZeeman}) of the $5D_{3/2}$ and $6S_{1/2}$ manifolds and equal initial state population across the $5D_{3/2}$ manifold.
        \textbf{d}, relative drift of the centre frequencies ($\sigma_{\Delta\omega}=2\pi\times10$~MHz) of the two photons over the course of the experiment due to laser drift.
        In all cases the experimental error bars denote statistical uncertainties.
        Additionally in \textbf{a}, we show both the interfering and non-interfering theory curves where we assume the photons are transform limited and there are no other experimental imperfections.
        All theory curves are normalised such that the maximum of the equivalent non-interfering curve is unity.
        For theory curves we account for experimental SPAD gating by integrating $t_0$ over the region the gate is open.
        }
    \label{fig:combo_time_resolve_theory}
\end{figure*}

\noindent
Under the assumption that the photon is transform limited we can write:

\begin{equation}
    \zeta_i(t) = a_i(t) e^{-i \omega_i t} ,
\end{equation}

\noindent
where $a_i(t)$ is given by the temporal envelope of the photon and $\omega_i$ the centre frequency.
Without loss of generality, we assume $a_i(t)\in\mathbb{R} \ , \forall t$, then:

\begin{equation}
    \begin{split}
        P(t_0,\tau) &= \frac{1}{4} \left[ a_{\text{atom}}^2(t_0)a_{\text{ion}}^2(t_0+\tau) + a_{\text{atom}}^2(t_0+\tau)a_{\text{ion}}^2(t_0) \right. \\ 
        &\qquad\left. - 2\cos(\Delta \omega \ \tau)a_{\text{atom}}(t_0)a_{\text{atom}}(t_0+\tau)a_{\text{ion}}(t_0)a_{\text{ion}}(t_0+\tau) \right],
    \end{split}
    \label{prob}
\end{equation}

\noindent
where $\Delta\omega=\omega_{\text{ion}}-\omega_{\text{atom}}$ is the centre frequency difference between the two photons.
In general we are interested in the coincidence profile as a function of the relative detection time on two detectors, as in the main text.
This is determined from equation (\ref{prob}) by integrating over $t_0$:

\begin{equation}
    \mathcal{C}_{\text{theory}} (\tau) \propto \int dt_0 \, P(t_0,\tau) ,
    \label{coinc_from_P}
\end{equation}

\noindent
where the integral is taken with limits such that any temporal gating is accounted for.
For the remainder of this work we take the integral to be over a region that encompasses $\approx80$\% of the area of the ion-produced photon, to reproduce the effect of the gating of the SPAD in the experiment.
It is this gating which is responsible for the asymmetric shape seen in the experimental coincidences and the theoretical curves shown in Fig.~\ref{fig:combo_time_resolve_theory}.

Using equation (\ref{prob}) and (\ref{coinc_from_P}), along with the profiles for the two photons shown in Fig.~\ref{fig:od_g2_hom}a of the main text, we compare the expected and observed shapes of the HOM dip, shown in Fig.~\ref{fig:combo_time_resolve_theory}a, assuming the photons are transform limited and have identical centre frequencies.
To obtain the curve where the two photons do not interfere we note that in the limit $\Delta\omega\rightarrow\infty$ the final term in equation (\ref{prob}) oscillates rapidly as a function of $\tau$ and will average to zero with a finite detection bandwidth.
A qualitative discrepancy is seen between the experimental observations and what is theoretically expected in the case where $\Delta\omega=0$.
However, the experimental data, outside of the dip around $\tau\approx0$, matches what we would expect for non-interfering photons.
This indicates that we are over-estimating the width of the HOM dip.

\newpage
There are several explanations why the experimentally observed dip is narrower than that expected by theory discussed above.
The theory used in Fig. \ref{fig:combo_time_resolve_theory}a assumes $\Delta\omega=0$ and that the single photon pulses are transform limited.
These assumptions are broken in the experiment in the following ways:
\begin{itemize}
    \item for the ion, decay from the $6P_{1/2}$ manifold to the $5D_{3/2}$ manifold and subsequent re-exciation during the extraction phase destroys the transform limitedness of the produced photons
    \item experimental uncertainties in the detunings of the 493-nm and 650-nm laser frequencies from their corresponding resonance frequencies causes uncertainties in $\Delta\omega$
    \item drift on the ion laser locks produces a corresponding drift in $\Delta \omega$
    \item the scheme used to pump the ion into the  $5D_{3/2}$ manifold, as well as  Zeeman splitting, causes the ion to emit photons at multiple frequencies
\end{itemize}
In the following, we explore the effect of the breaking of these assumptions.

Following the initial excitation from the $5D_{3/2}$ manifold to the $6P_{1/2}$ manifold, the ion can decay to either to the ground state with $\approx75\%$ probability, emitting a 493-nm photon, or back to the $5D_{3/2}$ manifold with $\approx25\%$ probability. 
If the latter occurs, re-excitation to the $6P_{1/2}$ manifold will result in emission of a 493-nm photon delayed relative to the photon that would have been emitted if the decay to the $5D_{3/2}$ manifold did not occur.  This creates a temporally lengthened observed photon shape but leaves the spectrum unchanged~\cite{EschnerQuantumEmitters}, resulting in a non-transform limited average temporal profile.
To account for this we calculate the transform limited pulse shape for the ion-produced photons by numerically solving optical Bloch equations.
To determine the expected coincidence curve when no scatter back to the $5D_{3/2}$ state occurs, we use equation (\ref{prob}) along with the ion's transform-limited pulse shape.
For the case where a single scattering event back to the $5D_{3/2}$ state occurs, equation (\ref{prob}) is again used along with the ion-produced photon's transform-limited profile, but here we add a probabilistic temporal displacement to account for the photons delayed emission.
The total non-transform limited theory curve, shown in Fig.~\ref{fig:combo_time_resolve_theory}a, is the sum of these two coincidence curves weighted by the branching ratio.
We do not account for higher order processes, where the ion scatters back to the $5D_{3/2}$ state more than once.
As shown by Fig. \ref{fig:combo_time_resolve_theory}a, taking into account the non-transform limitedness of the ion-produced photon marginally alters the expected coincidence curve from that where we assume the two photons to be transform limited.

While an effort was made to ensure that the photons produced by the two sources had identical centre frequencies, there were uncertainties in the actual value of $\Delta\omega$, predominantly due to the limited resolution of the ion spectroscopy and drifts in the ion laser locks.
From equation (\ref{prob}) it can be seen that a non-zero value of $\Delta\omega$ gives rise to an oscillating envelope to the final term, which can cause a reduction in the width of the HOM dip, as seen in Fig.~\ref{fig:combo_time_resolve_theory}b.

\begin{figure}
    \centering
    \includegraphics[width=0.9\textwidth]{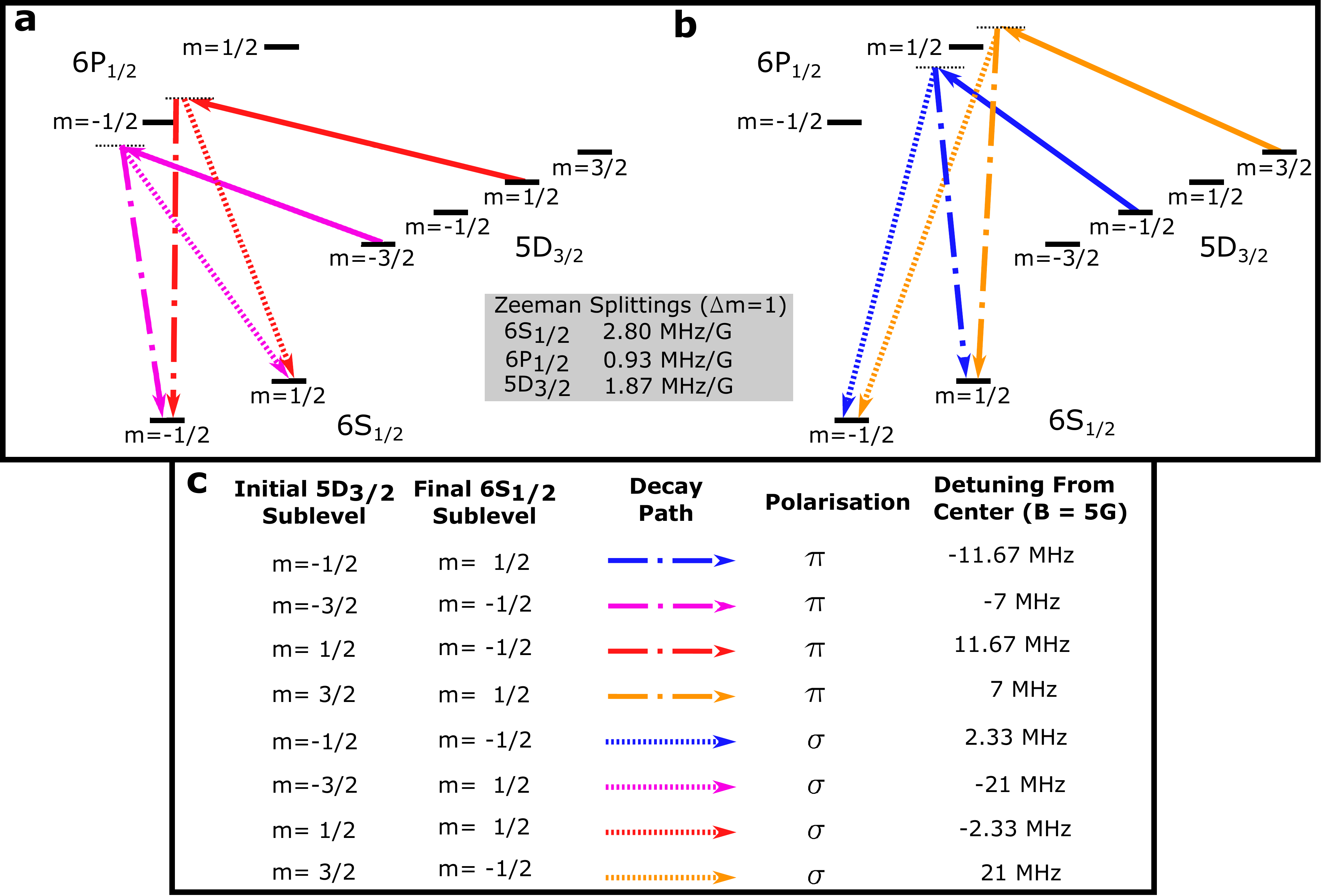}
    \caption{\textbf{Barium Photon Frequencies Including Zeeman Splittings.}
    In our excitation scheme, all possible Zeeman sublevels of $5D_{3/2}$ are equally populated, and only $\sigma$ excitation transitions are used. 
    \textbf{a} Possible excitation and decay paths involving the $\ket{6P_{1/2},m=-1/2}$ state of $^{138}$Ba$^+$. \textbf{b}, possible excitation and decay paths involving the $\ket{6P_{1/2},m=1/2}$ state of $^{138}$Ba$^+$. \textbf{c}, shows the polarisation and detuning of the resulting 493-nm photons relative to the centre of all possible emission frequencies.}
    \label{fig:BaZeeman}
\end{figure}

Throughout the experiment the ion was subjected to a magnetic bias field ($\approx 5$ Gauss), lifting the degeneracy of the Zeeman sub-levels for the three states shown in Fig.~\ref{fig:layout}b of the main text.
Additionally, the ion is pumped such that the Zeeman states of the $5D_{3/2}$ are populated equally.
Thus the $493$-nm photons were emitted with a frequency shift given by the differential shift between the initial and final Zeeman sub-levels.
While the spectrum of a single ion-produced photon is practically monochromatic with a narrow spectral bandwidth, the average spectrum of the ion source consists of several spectral peaks separated by these differential shifts and weighted by their likelihood.
To account for this we modify equation (\ref{prob}):

\begin{equation}
    \begin{split}
        P(t_0,\tau) = &\frac{1}{4}\left[ a_{\text{atom}}^2(t_0)a_{\text{ion}}^2(t_0+\tau) + a_{\text{atom}}^2(t_0+\tau)a_{\text{ion}}^2(t_0)\right.\\ 
        &\qquad\left. - 2 a_{\text{atom}}(t_0)a_{\text{atom}}(t_0+\tau)a_{\text{ion}}(t_0)a_{\text{ion}}(t_0+\tau) \sum_i c_i \cos((\Delta\omega_i+\Delta\omega) \tau) \right],
    \end{split}
    \label{prob_teeth}
\end{equation}

\noindent
where $c_i$ and $\Delta\omega_i$ are the weighting and differential shifts due to Zeeman splitting, and $\sum_i c_i = 1$.
Given the population spread across the $5D_{3/2}$ sub-levels, as well as the polarisation and propagation direction of the 650-nm excitation light, we expect photons to be emitted from the ion at several frequencies around a mean value with near-equal probability, shown in Fig.~\ref{fig:BaZeeman}.
As seen in Fig.~\ref{fig:combo_time_resolve_theory}c, this type of probabilistic spectrum gives rise to a narrowed HOM dip with subsequent oscillations in coincidence space appearing less pronounced.

\begin{figure*}[h]
    \centering
    \includegraphics[width=0.75\textwidth]{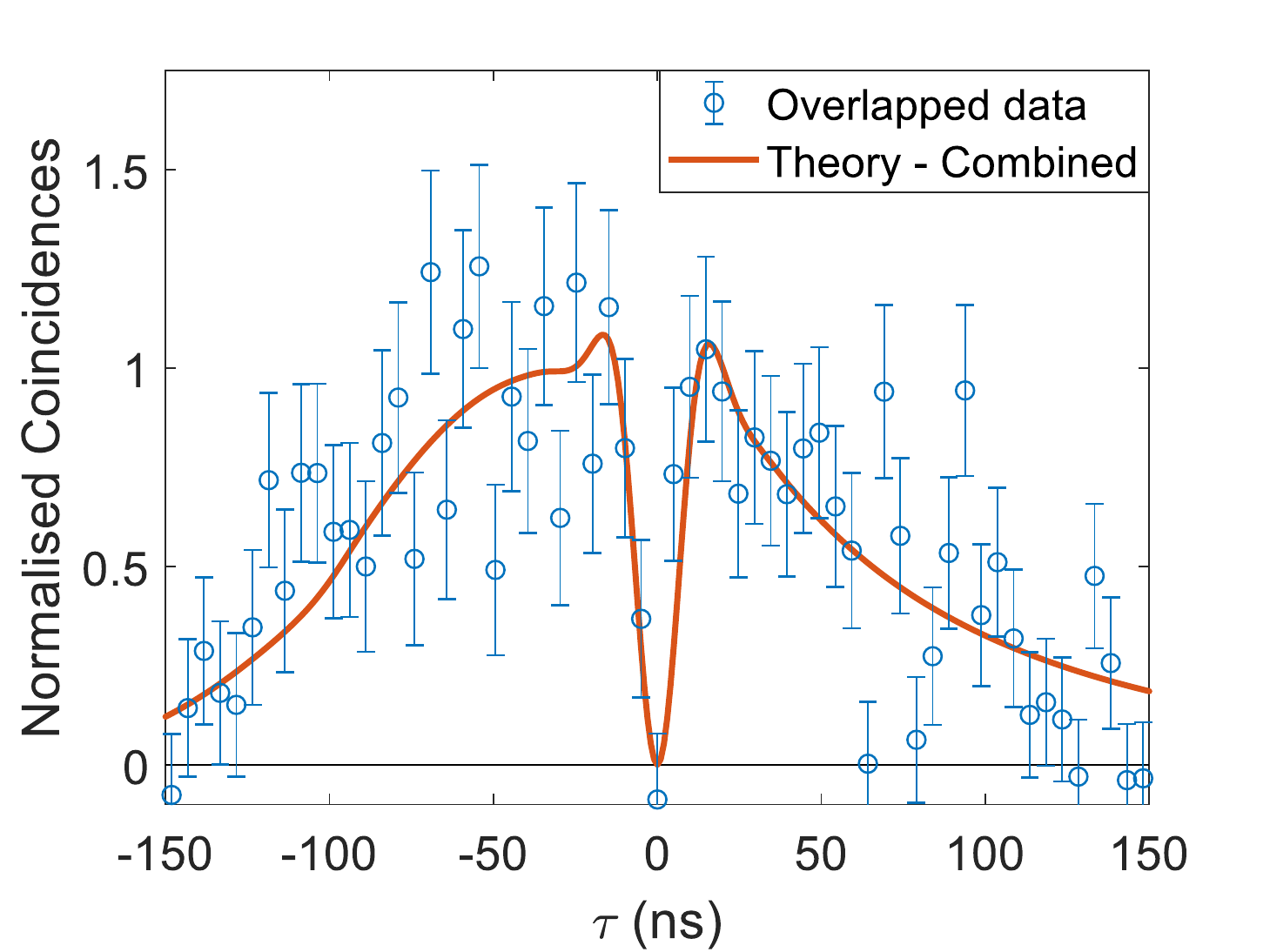}
        \caption{\textbf{Theory accounting for combination of experimental imperfections.}
        Experimental data (blue circles) and the theory curve (orange) calculated from equation (\ref{tot_time_resolved_theory}) showing the combination of the separate experimental imperfections described in Fig.~\ref{fig:combo_time_resolve_theory}. The error bars on the experimental data denote statistical uncertainties}
    \label{fig:tot_time_resolve_theory}
\end{figure*}

Due to the way the DFG light was frequency stabilised, it is likely that the ion produced photon frequency drifted relative to that of the atomic-ensemble produced photon.
To include this drift in the theory we modify equation~\ref{prob}:

\begin{equation}
    \begin{split}
        P(t_0,\tau) &= \int_{-\infty}^\infty d\Delta\omega \frac{e^{-\frac{(\Delta\omega-\Delta\omega_0)^2}{2\sigma_{\Delta\omega}^2}}}{\sigma_{\Delta\omega}\sqrt{2\pi}} P(t_0,\tau)\\
        &= \frac{1}{4}\left[ a_{\text{atom}}^2(t_0)a_{\text{ion}}^2(t_0+\tau) + a_{\text{atom}}^2(t_0+\tau)a_{\text{ion}}^2(t_0)\right.\\
        &\qquad\left.- 2\cos(\Delta\omega_0\tau)e^{-\frac{1}{2} \sigma_{\Delta\omega}^2\tau^2} a_{\text{atom}}(t_0)a_{\text{atom}}(t_0+\tau)a_{\text{ion}}(t_0)a_{\text{ion}}(t_0+\tau) \right],
    \end{split}
    \label{prob_jitter}
\end{equation}

\noindent
where we've assumed a Gaussian profile to the drift with an average detuning, $\Delta\omega_0$, and variance, $\sigma_{\Delta\omega}$.
Using values, $\Delta\omega_0 = 0$ and $\sigma_{\Delta\omega}=2\pi\times10$~MHz in Fig.~\ref{fig:combo_time_resolve_theory}d, we see that the theoretical HOM dip narrows while not exhibiting the large oscillations characteristic of a static frequency offset between the two photons.

To account for all the effects discussed above, equation (\ref{prob}) becomes:

\begin{equation}
    \begin{split}
        P(t_0,\tau) &= \frac{1}{4}\left[ a_{\text{atom}}^2(t_0)a_{\text{ion}}^2(t_0+\tau) + a_{\text{atom}}^2(t_0+\tau)a_{\text{ion}}^2(t_0)\right.\\
        &\qquad\left.\ - 2\mathcal{A}(t_0,\tau)\sum_i c_i \cos{(\Delta\omega_i + \Delta\omega_0)\tau} e^{-\frac{1}{2} \sigma_{\Delta\omega}^2\tau^2}\right],
    \end{split}
    \label{tot_time_resolved_theory}
\end{equation}

\noindent
where we have modified the original $a_{\text{atom}}(t_0)a_{\text{atom}}(t_0+\tau)a_{\text{ion}}(t_0)a_{\text{ion}}(t_0+\tau)$ term from equation (\ref{prob}) to include the non-transform limitedness of the ion produced photon, denoted by  $\mathcal{A}(t_0,\tau)$.
We use equation~(\ref{tot_time_resolved_theory}), with the values used for the individual plots in Fig.~\ref{fig:combo_time_resolve_theory}, to produce the theory curve in Fig.~\ref{fig:tot_time_resolve_theory} and those in Fig.~\ref{fig:od_g2_hom}b and c of the main text.
The factors discussed which pose the largest problems, namely effects which cause the centre frequencies of the two sources to be different either through experimental drift, constant offsets or differential Zeeman shifts, are correctable.
Therefore, there is scope to increase the HOM dip width, and thus increase entanglement rates by correcting these issues.

\subsection{Expected Fidelity of Hybrid Entanglement}\label{sec:expected_fidelity}

\begin{figure}
    \centering
    \includegraphics[width=1\textwidth]{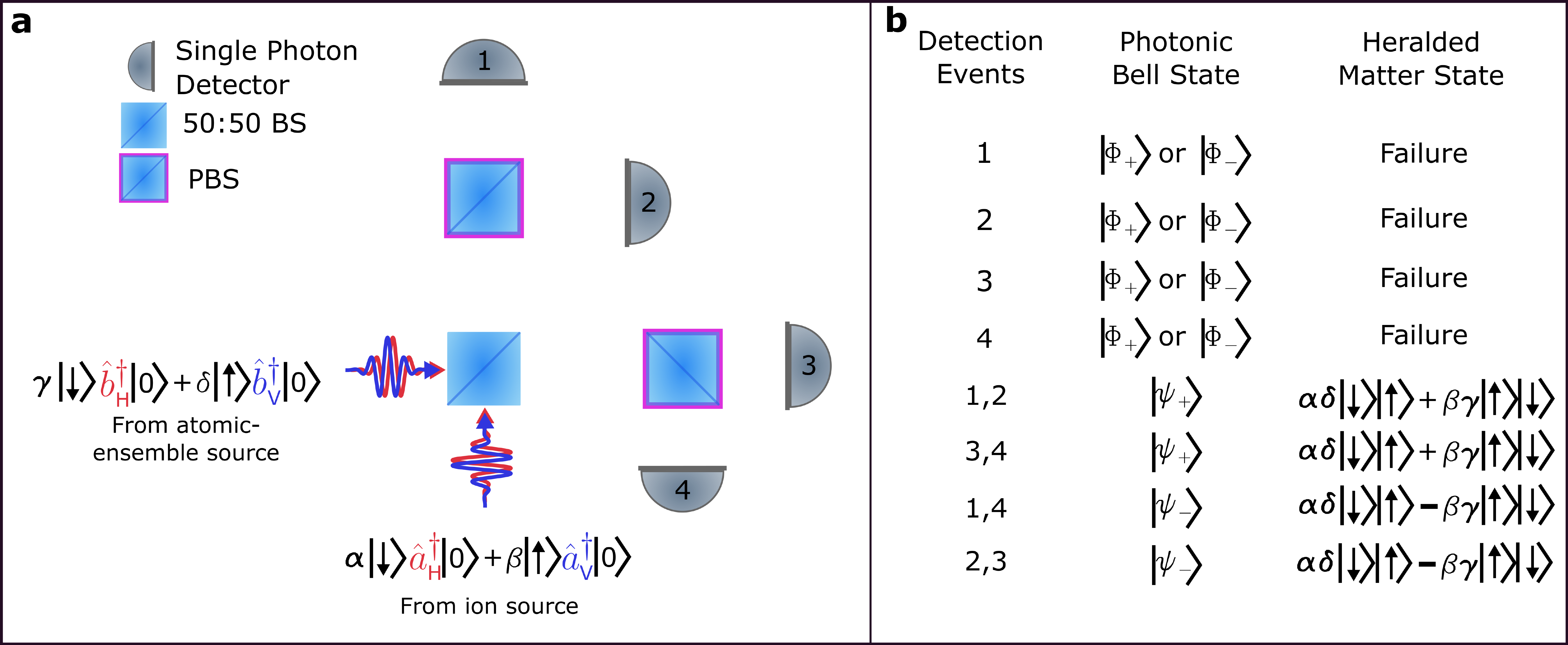}
    \caption{\textbf{Setup to herald entanglement between distance matter qubits.} \textbf{a}, incoming photons have their polarisations entangled with their corresponding matter qubit's internal states. a 50:50 beamsplitter (BS) is used to interfere the two photons, allowing for the heralding of entanglement between the matter qubits after detection using polarising beamsplitters (PBSs) and single photon detectors. \textbf{b}, different combinatinations of detector clicks correspond to the detection of certain photonic bell states. Depending on which set of detectors click (labeled in \textbf{a}), different entangled states between the matter qubits can be heralded. In the case of any individual detector clicking, the $\Phi_+$ and $\Phi_-$ photonic bell states cannot be distinguished from one another, resulting in a failed attempt to entangle the matter qubits. All other combinations of detector clicks not shown should not be possible in the case of perfect two-photon interference, and are ignored in the case of imperfect interference.}
    \label{fig:entanglement}
\end{figure}

We consider a common entanglement generation scheme, described in \cite{simon2003robustentanglement}, which uses a setup consisting of a 50:50 beamsplitter (BS), followed by two polarising beamsplitter (PBS) and four single-photon detectors, as shown in Fig. \ref{fig:entanglement}. 
For the purposes of general analysis, we will assume that the ion and atomic ensemble produce single photons, where the polarisation degree of freedom of each photon is entangled with some internal state of its source system. 
With the photons from each source arriving at the beamsplitter simultaneously, we write the total state of the system as:

\begin{equation}
    \begin{split}
         \ket{\Psi_1}_{\text{in}}&= (\alpha\ket{\downarrow}\ket{H}_{\text{in}}+\beta\ket{\uparrow}\ket{V}_{\text{in}})\otimes(\gamma\ket{\downarrow}\ket{H}_{\text{in}}+\delta\ket{\uparrow}\ket{V}_{\text{in}})\\
         &=\alpha\gamma\ket{\downarrow\downarrow}\ket{H,H}_{\text{in}}+\beta\delta\ket{\uparrow\uparrow}\ket{V,V}_{\text{in}}+\alpha\delta\ket{\downarrow\uparrow}\ket{H,V}_{\text{in}}+\beta\gamma\ket{\uparrow\downarrow}\ket{V,H}_{\text{in}},
    \end{split}
\end{equation}

\noindent
where we have made the assumption that horizontal (vertical) photon states, denoted by $\ket{H}(\ket{V})$, are entangled with the $\ket{\downarrow}$($\ket{\uparrow}$) internal states of their corresponding atomic systems. Here $\alpha$, $\beta$, $\gamma$ and $\delta$ are the quantum mechanical  probability amplitudes of their respective states. It is informative to rewrite $\ket{\Psi_1}_{\text{in}}$ in the photonic Bell basis  \cite{QuantumLogicTrappedIons}:

\begin{equation}\label{inputbell}
    \begin{split}
        \ket{\Psi_{1}}_{\text{in}}&=\alpha\gamma\ket{\downarrow\downarrow}(\ket{\Phi_+}_{\text{in}}+\ket{\Phi_-}_{\text{in}})+\beta\delta\ket{\uparrow\uparrow}(\ket{\Phi_+}_{\text{in}}-\ket{\Phi_-}_{\text{in}})\\
        &\qquad+\alpha\delta\ket{\uparrow\downarrow}(\ket{\psi_+}_{\text{in}}+\ket{\psi_-}_{\text{in}})+\beta\gamma\ket{\downarrow\uparrow}(\ket{\psi_+}_{\text{in}}-\ket{\psi_-}_{\text{in}})\\
        &=(\alpha\gamma\ket{\downarrow\downarrow}+\beta\delta\ket{\uparrow\uparrow})\ket{\Phi_+}_{\text{in}}+(\alpha\gamma\ket{\downarrow\downarrow}-\beta\delta\ket{\uparrow\uparrow})\ket{\Phi_-}_{\text{in}}\\
        &\qquad+(\alpha\delta\ket{\downarrow\uparrow}+\beta\gamma\ket{\uparrow\downarrow})\ket{\psi_+}_{\text{in}}+(\alpha\delta\ket{\downarrow\uparrow}-\beta\gamma\ket{\uparrow\downarrow})\ket{\psi_-}_{\text{in}},
    \end{split}
\end{equation}
\noindent

where

\begin{equation}
    \begin{split}
        \ket{\Phi_{\pm}}&=\frac{1}{\sqrt{2}}(\ket{H,H}\pm\ket{V,V})\\
        \ket{\psi_{\pm}}&=\frac{1}{\sqrt{2}}(\ket{H,V}\pm\ket{V,H}).\\
    \end{split}
\end{equation}

In the following, we use similar notation to section \ref{sec:visDeriv} for the definitions of $\hat{a}^{\dagger}$,$\hat{b}^{\dagger}$,$\hat{x}^{\dagger}$,$\hat{y}^{\dagger}$, with additional subscripts of H and V denoting horizontally and vertically polarised photons respectively.
Computing the action of the beamsplitter on the $\ket{\Phi_{\pm}}$ photonic Bell states, one can show \cite{QuantumLogicTrappedIons}:

\begin{equation}
    \label{phibell}
    \begin{split}
        \ket{\Phi_{\pm}}_{\text{in}}&=\frac{1}{\sqrt{2}}(\ket{H,H}_{\text{in}}\pm\ket{V,V}_{\text{in}})=\frac{1}{\sqrt{2}}(\hat{a}_H^{\dagger}\hat{b}_H^{\dagger}\pm\hat{a}_V^{\dagger}\hat{b}_V^{\dagger})\ket{0,0}_{\text{in}}\\
        &\rightarrow\frac{1}{2\sqrt{2}}[(\hat{x}_H^{\dagger}+i\hat{y}_H^{\dagger})(i\hat{x}_H^{\dagger}+\hat{y}_H^{\dagger})\pm(\hat{x}_V^{\dagger}+i\hat{y}_V^{\dagger})(i\hat{x}_V^{\dagger}+\hat{y}_V^{\dagger})]\ket{0,0}_{\text{out}}\\
        &\rightarrow\frac{i}{2}\left[\ket{HH,0}_{\text{out}}+\ket{0,HH}_{\text{out}} \pm  \left(\ket{VV,0}_{\text{out}} + \ket{0,VV}_{\text{out}}\right)\right],
    \end{split}
\end{equation}
where notation such as $\ket{ij,0}$ corresponds two photons with polarisation $i$ and $j$ exiting the same port of the 50:50 beamsplitter. 
In a similar fashion:
\begin{equation}
        \ket{\psi_+}_{\text{in}}\rightarrow\frac{i}{\sqrt2}(\ket{HV,0}_{\text{out}}+\ket{0,HV}_{\text{out}}),
\end{equation}
\begin{equation}
        \ket{\psi_-}_{\text{in}}\rightarrow\frac{1}{\sqrt2}(\ket{H,V}_{\text{out}}-\ket{V,H}_{\text{out}}).
\end{equation}

From eqn \ref{phibell} we see that the action of the beamsplitter on the $\ket{\Phi_+}_{\text{in}}$ and $\ket{\Phi_-}_{\text{in}}$ photonic Bell states results in photons with the same polarisation exiting out of the same ports of the 50:50 BS. Therefore, measurement using the setup shown in Fig. \ref{fig:entanglement} cannot distinguish $\ket{\Phi_+}_{\text{in}}$ from $\ket{\Phi_-}_{\text{in}}$. 
The $\ket{\psi_+}_{\text{in}}$ input state results in two oppositely polarised photons exiting the same port of the 50:50 BS, and the $\ket{\psi_-}_{\text{in}}$ input state results in two oppositely polarised photons exiting opposite ports of the 50:50 BS.
The scheme in  Fig.~\ref{fig:entanglement} can therefore distinguish between $\ket{\psi_+}_{\text{in}}$ and $\ket{\psi_-}_{\text{in}}$, and can be used to herald the desired ensemble-ion entangled state after detection of the two photons.

Thus far, we have assumed that the two-photon interference is perfect.
In general this is not the case, and to account any imperfection we use a parameter $c$, the mode overlap of the photons, not including polarisation.
We rewrite the input state as:

\begin{equation}
    \begin{split}
        \ket{\Psi}_{\text{in}}&=\left[\alpha\ket{\downarrow}(\sqrt{c}\ket{H}_{\text{in}}+\sqrt{1-c}\ket{H_n}_{\text{in}})+\beta\ket{\uparrow}(\sqrt{c}\ket{V}_{\text{in}}+\sqrt{1-c}\ket{V_n}_{\text{in}}\right]\otimes\left[\gamma\ket{\downarrow}\ket{H}_{\text{in}}+\delta\ket{\uparrow}\ket{V}_{\text{in}}\right]\\
        &=\sqrt{c}\ket{\Psi_1}_{\text{in}}+\sqrt{1-c}\ket{\Psi_2}_{\text{in}},
    \end{split}
\end{equation}
\noindent
where $\ket{H_{n}}$ and $\ket{V_{n}}$ represent polarised photons that are nonidentical to $\ket{H}$ and $\ket{V}$ in their spectral, spatial, or temporal profiles and $\ket{\Psi_2}_{\text{in}} = \alpha\gamma\ket{\downarrow\downarrow}\ket{H_n,H}_{\text{in}}+\beta\delta\ket{\uparrow\uparrow}\ket{V_n,V}_{\text{in}}+\alpha\delta\ket{\downarrow\uparrow}\ket{H_n,V}_{\text{in}}+\beta\gamma\ket{\uparrow\downarrow}\ket{V_n,H}_{\text{in}}$. The action of the beamsplitter on the first term has already been examined.
The action of the beamsplitter on the second term can be shown to give:

\begin{equation}
    \begin{split}
        \ket{\Psi_2}_{\text{in}}
        &\rightarrow\alpha\gamma\ket{\downarrow\downarrow}(i\ket{HH_n,0}_{\text{out}}+i\ket{0,HH_n}_{\text{out}}-\ket{H,H_n}_{\text{out}}+\ket{H_n,H}_{\text{out}})\\
        &+ \beta\delta\ket{\uparrow\uparrow}(i\ket{VV_n,00}_{\text{out}}+i\ket{0,VV_n}_{\text{out}}-\ket{V,V_n}_{\text{out}}+\ket{V_n,V}_{\text{out}})\\
        &+ \beta\gamma\ket{\uparrow\downarrow}(i\ket{HV_n,0}_{\text{out}}+i\ket{0,HV_n}_{\text{out}}+\ket{V_n,H}_{\text{out}}-\ket{H,V_n}_{\text{out}})\\
        &+ \alpha\delta\ket{\downarrow\uparrow}(i\ket{H_n V, 0}_{\text{out}}+i\ket{0, H_n V}_{\text{out}} - \ket{V,H_n}_{\text{out}}+\ket{H_n,V}_{\text{out}}).
    \end{split}
\end{equation}

Now we can investigate how the total output state is affected by the measurement of a coincidence between the detectors shown in Fig. \ref{fig:entanglement}.
As an example we use the measurement of the $\ket{\psi_{-}}_{\text{in}}$ photonic Bell state, which corresponds to the measurement of a horizontal and vertical photon out of opposite ports of the 50:50 beamsplitter. 
In this case of perfect interference this will herald the $\ket{\psi_-}_m = (\ket{\uparrow\downarrow}-\ket{\downarrow\uparrow})$/$\sqrt{2}$ matter Bell state. This is represented by the measurement operator:

\begin{equation}
\begin{split}
    M_{\mathcal{C}}=&\ket{H,V}\bra{H,V}+\ket{V,H}\bra{V,H}+\ket{H_n,V}\bra{H_n,V}+\ket{V,H_n}\bra{V,H_n}\\
    &+\ket{V_n,H}\bra{V_n,H}+\ket{H,V_n}\bra{H,V_n}+\ket{H_n,V_n}\bra{H_n,V_n}+\ket{V_n,H_n}\bra{V_n,H_n}.\\
\end{split}
\end{equation}
\noindent
In the case of perfect interference, any terms containing $H_n$ or $V_n$ are unnecessary.
We calculate the density matrix describing state of the two matter qubits after such a measurement:

\begin{equation}
    \rho_{m}=Tr_{\text{photons}}[M_{\mathcal{C}}\ket{\Psi}_{\text{out}}\bra{\Psi}_{\text{out}}]=
    N \left(c\begin{bmatrix}
    0&0&0&0\\
    0&\abs{\alpha}^2\abs{\delta}^2&-\alpha\beta^*\gamma^*\delta&0\\
    0&\alpha^*\beta\gamma\delta^*&\abs{\beta}^2\abs{\gamma}^2&0\\
    0&0&0&0
    \end{bmatrix}
    +(1-c)
    \begin{bmatrix}
    0&0&0&0\\
    0&\abs{\alpha}^2\abs{\delta}^2&0&0\\
    0&0&\abs{\beta}^2\abs{\gamma}^2&0\\
    0&0&0&0
    \end{bmatrix}\right),
\end{equation}

\noindent
where N is a normalisation factor given by:

\begin{equation}
    N=\frac{1}{\abs{\alpha}^2\abs{\delta}^2+\abs{\beta}^2\abs{\gamma}^2},
\end{equation}

\noindent
and where $\ket{\Psi}_{\text{out}}$ represents the state $\ket{\Psi}_{\text{in}}$ after exiting the beamsplitter.
If we set $\alpha=\delta=\beta=\gamma=1/\sqrt{2}$ in $\ket{\Psi}_{\text{in}}$, our measurement will herald the matter state $\ket{\psi_-}_{m}$ with a fidelity:
\begin{equation}
    F=\ \bra{\psi_-}_{m}\rho_{m}\ket{\psi_-}_{m}=\frac{1+c}{2}.
\end{equation}
If both sources have $g^{(2)}(0)=0$, as is the case for the on-demand photons in the main text, $c$ is equal to the visibility of the two-photon interference as seen in equation (\ref{eqn:visibility_coal}). We then arrive at the equation given in the main text,
\begin{equation}
    F=\frac{1+V}{2}.
\end{equation}
We note that a similar analysis gives the same result for the heralding of the $\ket{\psi_+}_m$ Bell state.
Additionally this analysis is valid for other types of qubits, such as time-binned or frequency qubits, as long as the PBSs used here are replaced with the relevant measurements for the other types of qubits (for instance dichroic mirrors could be used for frequency qubits). 

\subsection{Projected Hybrid Entanglement Rates}

To calculate prospective entanglement rates, we take the total number of background-subtracted coincidences ($\approx40$ for 5-ns bins) at $\tau=0$ for the non-overlapped on-demand case, divided by the total experimental run time ($\approx 22$ hours). 
Using the scheme described in section \ref{sec:expected_fidelity} we would reduce our entanglement rates by a factor of two, as only half of the possible Bell states can be heralded. 
However, in our experiment only half of the ion-produced photons were used due to polarisation filtering. By using all of the photons collected from the ion, this factor of two can be recovered.

Table \ref{tab:FidelityAndRates} gives fidelities and the expected entanglement rates achievable with our current experimental configuration using various bin sizes for our coincidence measurements. 
As can be seen, one must trade off bin size (entanglement rate) for fidelity.
The final column of Table \ref{tab:FidelityAndRates} shows our projections for entanglement rates given improvements in difference frequency generation (DFG) efficiency ($10\% \rightarrow 30\% $~\cite{LanyonConversionEfficiency,eschnerfreqconv}), ion-produced photon collection efficiency (using 0.6 NA lens with $70\%$ fibre coupling~\cite{ghadimi2017} as opposed to 0.4 NA lens with $35\%$ coupling efficiency) improvements in single photon detector efficiency ($70\% \rightarrow 90\% $ at 780-nm, available with superconducting nanowire single photon detectors from, as examples, Single Quantum, Photon Spot, and Quantum Opus), removal of fibre butt coupling losses (factor of 1.4), and reduction of losses in the optics in the DFG filtering stage (factor of 1.5).

\begin{table}[!h]
    \centering
    \begin{tabular}{|c|c|c|c|}
        \hline
         Bin Size & Projected &Expected Entanglement Rate& Projected Entanglement \\
         & Fidelity & With Current Setup ($s^{-1}$)&With Experimental Upgrades ($s^{-1}$)\\
         \hline
         \hline 5 ns & 1.0 $\pm$ 0.1 & 5.1$\times 10^{-4}$ & 3.1$\times 10^{-2}$ \\
         \hline 10 ns & 0.93 $\pm$ 0.06 &1.2$\times 10^{-3}$& 7.5$\times 10^{-2}$ \\
         \hline 15 ns & 0.72 $\pm$ 0.06 &1.8$\times 10^{-3}$& 1.1$\times 10^{-1}$ \\
         \hline
    \end{tabular}
    \caption{\textbf{Fidelities and entanglement rates for different bin sizes.} 
    The bin size of 5 ns corresponds to that of the results presented in the main text. 
    Additionally, projections are shown for improvements in DFG conversion efficiency, photon collection, and optical losses.}
    \label{tab:FidelityAndRates}
\end{table}

\end{document}